\documentclass[aps,showpacs,showkeys] {revtex4}

\usepackage{mathrsfs}
\usepackage{graphicx}   
\usepackage{epsfig}
\newcommand{\e}{\mathrm{e}} 

\begin{document}

\title{Fast tuning of superconducting microwave cavities}


\author{M. Sandberg}
\author{C. M. Wilson}
\author{F. Persson}
\author{G. Johansson}
\author{V. Shumeiko}
\author{T. Bauch}
\author{P. Delsing}
\affiliation{Department of Microtechnology and Nanoscience,Chalmers University of Technology.}
\author{T. Duty}
\affiliation{University of Queensland, School of Physical Sciences, Brisbane, QLD  4072 Australia.}


\begin{abstract}
Photons are fundamental excitations of the electromagnetic field and can be captured in cavities. For a given cavity with a certain size, the fundamental mode has a fixed frequency {\it f} which gives the photons a specific "color". The cavity also has a typical lifetime $\tau$, which results in a finite linewidth $\delta${\it f}. If the size of the cavity is changed fast compared to $\tau$, and so  that the frequency change $\Delta${\it f} $\gg \delta${\it f}, then it is possible to change the "color" of the captured photons.
Here we demonstrate superconducting microwave cavities, with tunable effective lengths. The tuning is obtained by varying a Josephson inductance at one end of the cavity. We show data on four different samples and demonstrate tuning by several hundred linewidths in a time $\Delta t \ll \tau$.  Working in the few photon limit, we show that photons stored in the cavity at one frequency will leak out from the cavity with the new frequency after the detuning. The characteristics of the measured devices make them suitable for different applications such as dynamic coupling of qubits and parametric amplification.
\end{abstract}

\maketitle


\section{Introduction}
Superconducting transmission line resonators are useful in a number of applications ranging from X-ray photon detectors \cite{Day} to parametric amplifiers \cite{Haviland_parametric, Lehnert_APL, Yamamoto} and quantum
computation applications \cite {Wallraff_Nature, Majer_2007, sillanpaa}.  Very recently, there has been a lot of interest in tunable superconducting resonators \cite{Osborn, Haviland_parametric, Lehnert_APL, Sandberg, saclay_tune, Yamamoto}.  In these experiments the inductive properties of the superconductor or a Josephson junction is implemented as a tunable element and is tuned by a bias current or a magnetic field.
These devices have both large tuning ranges and high quality factors, we have recently shown \cite{Sandberg} that the speed at which these devices can be tuned is substantially faster than the lifetime of the cavity.

The interaction between a qubit and
superconducting coplanar waveguide (CPW) resonator can, due to the small mode volume, be very strong when they are resonant with each other\cite{Blais_PRA}.  However, the
interaction can be modulated, becoming weak when the qubit and the
cavity are off-resonance.

In 2004 Wallraff {\textit et al.}  \cite{Wallraff_Nature} demonstrated that a superconducting quantum bit, in the form of a Single Cooper pair Box (SCB), could be strongly coupled to a transmission line resonator with a high quality factor. This demonstration opened up a new field of physic known as circuit Quantum Electrodynamics (cQED), it also gave new possibilities of coupling superconducting quantum bits.

The theoretical aspects of using a tunable transmission line resonator for coupling of quantum bits was  investigated by Wallquist {\textit et al.} \cite{Wallquist_PHD, Wallquist_PRB}. It was concluded that such a device, with proper characteristics, can be used for dynamic coupling of quantum bits and a protocol for a controlled phase gate was also presented. In this paper fabrication and characterization of a fast tunable superconducting transmission line resonator for the purpose of qubit coupling is described. We show data for four different samples.

\section{The tunable cavity \label{sec:tune}}
\begin{figure}[tb]
\centering
\includegraphics[width=0.7\textwidth]{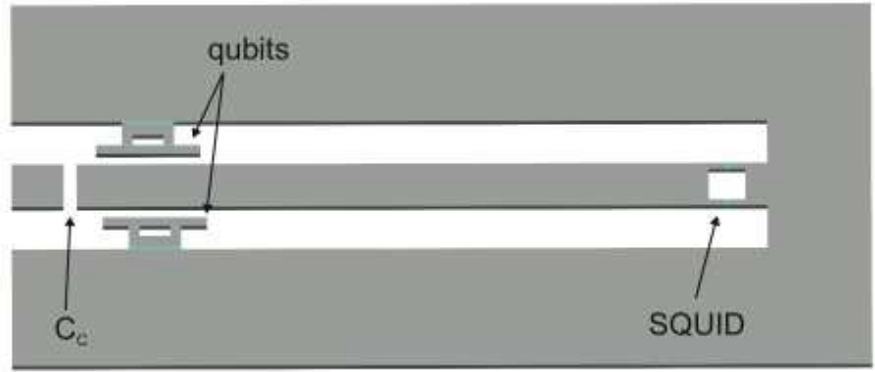}
\caption{\label{fig:circuitQED_SQUID}  A quarter wavelength resonator made tunable by inserting a SQUID in one end. The boundary condition at the SQUID end can be tuned by applying an external magnetic field. Two qubits are coupled capacitively to the open end of the cavity.}
\end{figure}

The possibility to achieve strong coupling between qubits and a cavity
makes circuit quantum electrodynamics a strong candidate for building a quantum information processor. In the experiment by Majer \textit{et al.} \cite{Majer_2007} where qubit coupling through a cavity was demonstrated and the experiment by Sillanp\"a\"a \textit{et al.} \cite{sillanpaa} where state transfer through a cavity was demonstrated the resonance frequency of the cavity was kept fixed while the eigenfrequencies of the qubits could be tuned. The need for fast individual control of each qubit, which can become quite complex with many qubits, could be overcome by using a tunable cavity (see figure\,\ref{fig:circuitQED_SQUID}). There are several different approaches possible for making the resonance frequency of the cavity tunable. Either a characteristic parameter of the transmission line could be varied \textit{i.e.} the inductance or capacitance per unit length by changing some property of the dielectric or the conductor, or by changing the boundary conditions of the cavity. Here we will focus on tunability obtained by changing the boundary condition of the cavity.

Instead of using a full wavelength resonator as in experiment by Wallraff \textit{et al.}\cite{Wallraff_Nature} a quarter wavelength resonator can be used. A quarter wavelength resonator, unlike the full wavelength resonator, is grounded in one end, at the other end the transmission line is open. Due to these boundary conditions resonance occurs for signals with a wavelength $\lambda$ such that   $\lambda=4\ell/(2n+1)$ where $\ell$ is the length of the transmission line and $n$ is a non-negative integer.

To make the quarter wavelength resonator tunable the short circuit at one end can be replaced by a tunable impedance. Since the circuit should behave quantum mechanically for the purpose of circuit-QED the tunable impedance must not introduce any large dissipation to the system. One possible tunable impedance fulfilling this criteria is the tunable inductance of a SQUID. The resonator, showed in figure\,\ref{fig:circuitQED_SQUID}, can now be tuned by applying an external magnetic flux to the SQUID loop. 

Under the conditions of low temperature and small dissipation electrical circuits can behave quantum
mechanically. The quantum mechanical Hamiltonian for such circuits can be
obtained starting from the classical Lagrangian. A Legendre transform of the Lagrangian and the imposing of
commutation relations gives the Hamiltonian \cite{Quantum_fluctuation}. The Lagrangian is usually obtained from the kinetic and potential energy of a system expressed in some generalized coordinates and their time derivatives. For an electrical circuit like this we use the capacitive and inductive energy of the components. 

\begin{figure}[tb]
\centering
\includegraphics[width=0.7\textwidth]{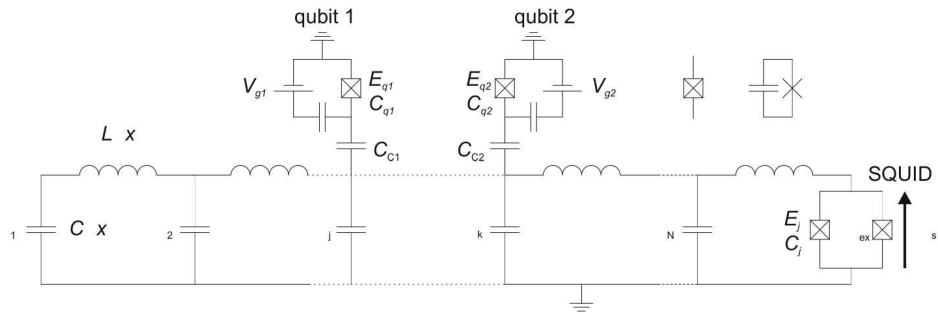}
\caption{\label{fig:circuitQED_SQUID_model}  b) Circuit diagram of quarter wavelength resonator coupled to two qubits. The boundary condition at the SQUID end can be tuned by applying an external magnetic field.}
\end{figure}

We start from the circuit schematic (ignoring the qubits), see figure\,\ref{fig:circuitQED_SQUID_model}, and using the procedure for quantization of  electrical circuits, following \cite{Wallquist_PHD, Wallquist_PRB}. Using the generalized flux,
$\Phi(x,t)=\int V dt$ to describe dynamics of the cavity, we get the Lagrangian and the equations of motion.  Solving these equations we see that the field inside the cavity must be of the form

\begin{equation}
\Phi(x,t)=\Phi_1\cos(kx)\sin \left( \frac{kt}{\sqrt{LC}} \right),
\end{equation}

where $C$ and $L$ are the capacitance and inductance per unit length of the transmission line, and where $k$ is the wave number of the field that has to be determined. Furthermore, we know have the boundary condition that $\Phi(\ell,t)=\Phi_s$, where $\Phi_s=(\Phi_{s1}+\Phi_{s2})/2$ is the average flux over the SQUID. 
Combining the bulk solution with the boundary condition and assuming small currents gives the dispersion relation

\begin{equation}
\label{eq:disp}
k\ell \tan k\ell = \frac{2\pi L\ell I_s}{\Phi_0} \left|\cos\frac{\pi\Phi_{ex}}{\Phi_0}\right|-\frac{C_s}{C\ell}(k\ell)^2,
\end{equation}

where $C_s$ is the sum capacitance of the SQUID junctions  and $\Phi_{ex}=(\Phi_{s1}-\Phi_{s2})$ is the magnetic flux in the SQUID loop. The SQUID is assumed to have identical junctions each with critical current $I_s/2$ and capacitance $C_s/2$. We can then get the wave number by numerically solving the dispersion equation.

\begin{figure}[t]
\centering
\includegraphics[width=0.7\textwidth]{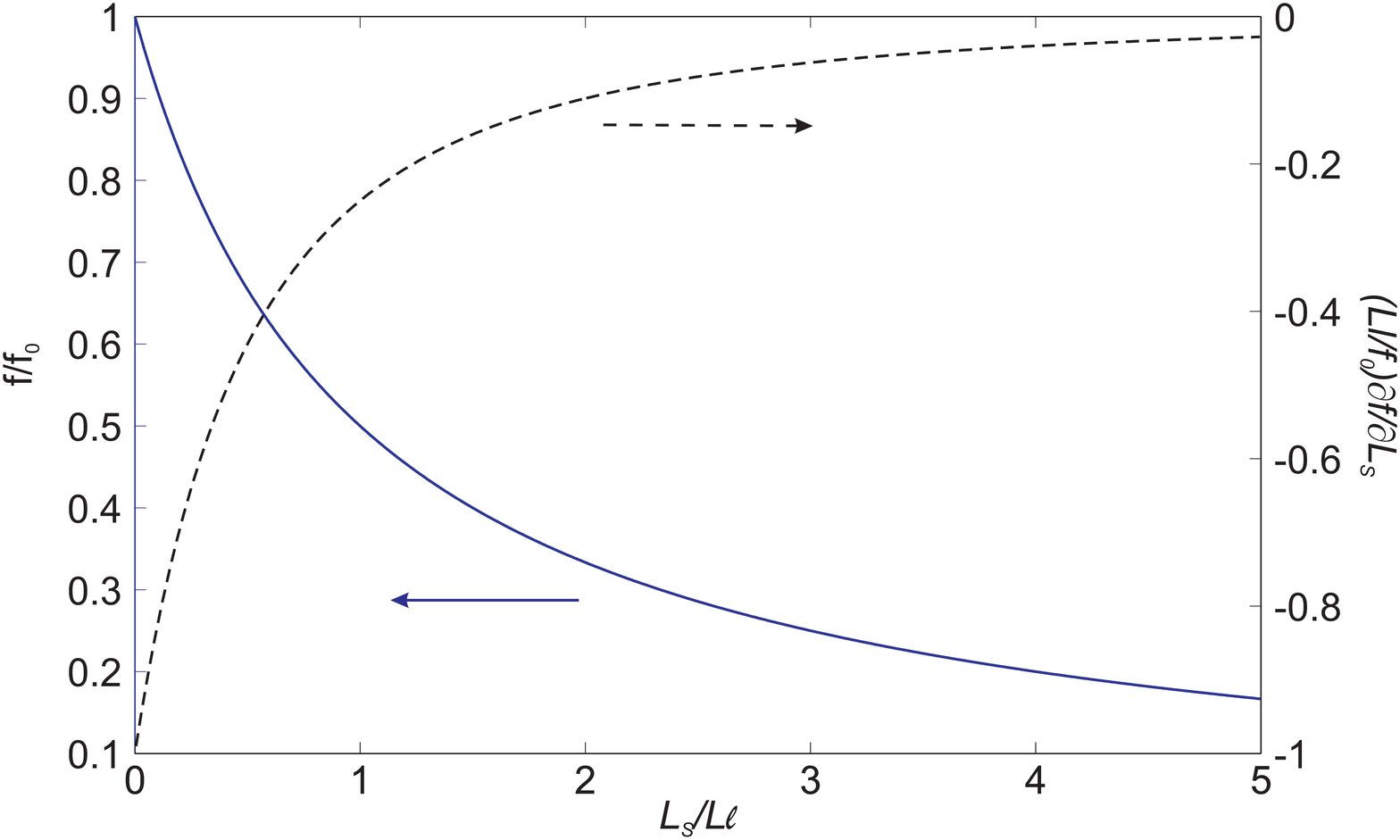}
\caption{\label{fig:tune1} Normalized frequency ($f/f_0$) (solid) and its derivative (dashed) as a function of the SQUID inductance $L_s$ normalized to the total cavity inductance $L\ell$.}
\end{figure}

A better understanding can be obtained by neglecting the last term in eq.\,\ref{eq:disp} by assuming that $C_s/C\ell$ is small, which is usually the case.  The dispersion relation can then be rewritten as

\begin{equation}
\frac{\cot k\ell}{k\ell} = \frac{\Phi_0}{2\pi L\ell I_s} \left| \cos \frac{\pi\Phi_{ex}}{\Phi_0}\right|
\end{equation}

expanding $\cot k\ell$ around $k\ell=\pi/2$ (corresponding to a infinite Josephson energy) and using $k=2\pi f\sqrt{LC}$ gives the resonance frequency $f$ as

\begin{equation}
\label{eq:resonace_f}
f=\frac{f_0}{1+L_s/L \ell}
\end{equation}

where $f_0$ is the quarter wavelength resonance frequency for a sample where the SQUID is replaced by a short, and

\begin{equation}
L_s=\frac{\Phi_0}{2 \pi I_s \left|\cos\frac{\pi\Phi_{ex}}{\Phi_0}\right|}
\end{equation}

is the inductance of the SQUID. In figure\,\ref{fig:tune1} the resonance frequency and its derivative as a function of SQUID inductance is plotted. By applying a magnetic field $\Phi_{ex}$ to the SQUID loop the inductance and hence the frequency of the cavity can be tuned. The qubits can now be addressed individually by the cavity if they have different eigenfrequencies. With this system Wallquist \textit{et al.} \cite{Wallquist_PRB} showed that fast dynamic coupling between qubits could be achieved.

\section{Sample Design}
\label{sec:design}
\begin{figure}
\centering
\includegraphics[width=0.7 \textwidth]{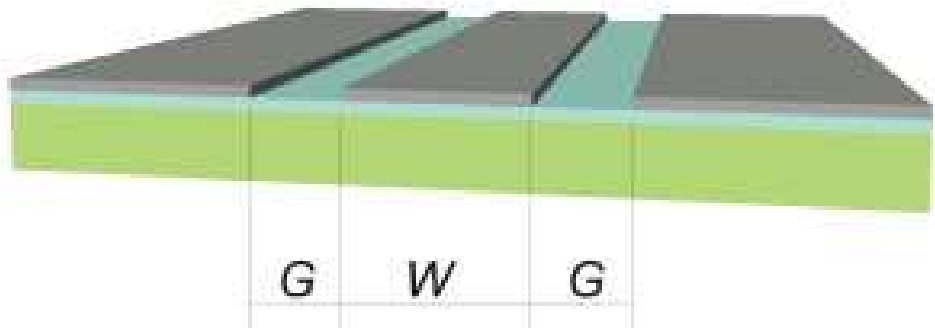}
\caption{\label{fig:CPW} The cavities are made from aluminum in a coplanar waveguide structure on top of an oxidized silicon  substrate. The gaps $G$ and the center strip $W$ and the substrate parameters determines the characteristic impedance of the transmission line.}
\end{figure}

We start be describing the transmission line which consists of a Co-Planar-Waveguide (CPW) with a center strip of width $W$ separated by a gap of width $G$ from a ground plane on each side, see figure\,\ref{fig:CPW}. The structure is fabricated on top of some dielectric material with a relative dielectric constant $\epsilon_r$. The characteristic impedance $Z_c$ of the transmission line depends on the center strip width, the gap width and the dielectric constant of the dielectric. To obtain $Z_c$ a conformal mapping technique can be used \cite{Collin}. Following the procedure of Gevorgian \textit{et al.} \cite{CAD} we can calculate $Z_c$, the capacitance per unit length $C$ and the effective dielectric constant $\epsilon_{eff}$. 

To be compatible with the SCB technology \cite{kevin_thesis} the aluminum is chosen for the resonator material and it is fabricated on a Si substrate with a wet grown insulating layer of SiO$_2$ with an effective dielectric constant $\epsilon_{eff} \approx 6.06$. Choosing the center strip width $W$ to be 13 $\mu$m and a gap width of $G=7 \mu$m, we obtain an impedance $Z_c$ = 50 $\Omega$, corresponding to a capacitance per unit length of 164\,\rm{pF/m}.

To be in the range of typical SCB eigenfrequencies the cavity frequency is designed to be $f\approx 5$\,GHz. The length of a quarter wavelength resonator is obtained as $\ell=c/4\sqrt{\epsilon_{eff}}f$, where $c$ is the speed of light in vacuum.

To probe the cavity it has to be coupled to the outside world, which is done via a coupling capacitance $C_c$. The coupling capacitance will also lead to leakage of energy out of the resonator and hence to a lower Q value. The (external) Q value for a resonator consisting of an inductance $L_r$ in parallel with a capacitance $C_r$ that is coupled to a load resistance $R$ through the capacitance $C_c$ is obtained as

\begin{equation}
\label{eq:Q_ext}
Q_{ext}=\frac{(1+(\omega_rC_cR)^2)(C_r+C_c)}{\omega_rC_c^2R}
\end{equation}

where $\omega_r=1/\sqrt{L_r(C_r+C_c)}$ is the resonance frequency. The same expression is obtained for a quarter wavelength resonator if the capacitance $C_r$ is replaced by $C\ell/2$. To obtain an external Q-value of $10^{4}$ for a 5\,GHz resonator coupled to a 50 $\Omega$ environment we would need a coupling capacitance of $C_c = 5.6 fF$. We have used an interdigitated capacitance with only one finger on each side. The approximate length, width and spacing of the fingers were obtained by using the microwave simulation program Microwave Office.

The internal $Q$ is can be limited by a number of different mechanisms. Here we list four different possible sources of dissipation that can contribute to the internal Q-value.
i) At high frequencies some of the electric field will penetrate into the superconductor. At finite temperatures, below the critical temperature $T_c$, there are still some unpaired electrons above the superconducting energy gap (quasi-particles). The quasi particles causes dissipation when they interact with the electric field. ii) When the cavity oscillates a high frequency current is passed through the SQUID inductance, which will also generate a voltage over the SQUID. This voltage can drive a current though the sub-gap resistance of the SQUID which causes dissipation. iii) There are also losses in the dielectric of the CPW which can lead to dissipation. This is typically described as a complex dielectric constant and a loss tangent. iv) Flux noise in the SQUID loop will cause fluctuations in the resonance frequency and would also result in a larger line width. The sensitivity to flux noise is determined by the derivative $\delta f/\delta \Phi$. We will return to what limits the internal Q-value.

Next we describe the tunable element which is included in the cavity, namely the Superconducting Quantum Interference device (SQUID).
When connecting a Josephson junction or a SQUID to the resonator one has to consider that the junction itself acts as a resonator with a resonance frequency $\omega_p$, called the plasma frequency. In order not to excite the SQUID, the resonance frequency of the cavity, $\omega_r$ must be much less than the SQUID plasma frequency, 
$\omega_ p=\sqrt{2\pi I_{s}/(\Phi_0 C_s)}$. This gives the requirement $I_s\gg \omega_r^2\Phi_0C_s/2\pi$.
The capacitance of Al/AlO$_x$/Al Josephson junctions is approximately 45 fF/$\mu$m$^2$ \cite{Delsing_Claeson_1990}. Assuming a 1$\mu$m$^2$ area this requires $I_s \gg 15 $$n$A.
One also has to make sure that the temperature $T$ is low enough so that $\hbar \omega_p \gg \hbar w_r \gg k_BT$.
This gives the restriction $T\ll 270$ mK on the temperature for a 5\,GHz resonator.

\begin{figure}
\centering
\includegraphics[width=0.7\textwidth]{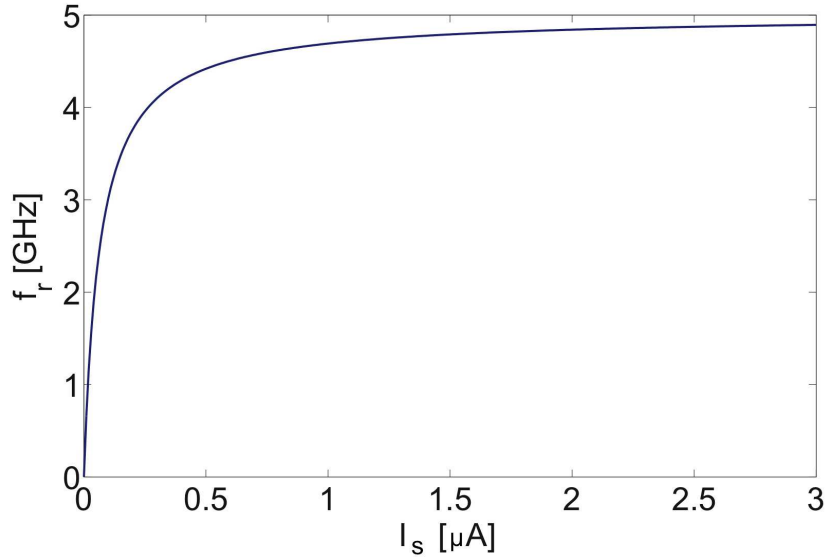}
\caption{\label{fig:f_vs_Is}Calculated resonance frequency $f_r$ as a function of the critical current $I_s$ of the SQUID using the design parameters for the cavity described in the text.}
\end{figure}

How much $I_s$ can be suppressed by the magnetic field depends on the symmetry of the junctions. The amount that $I_s$ can be suppressed depends on how identical the Josephson junctions can be fabricated. If we assume that they can be fabricated with an $I_c$ within 2\% then the minimum $I_s$  is going to be $\sim$$0.02\,I_c$. For the parameters obtained in the last section we see that we have most tunability for low $I_s$, see figure\,\ref{fig:f_vs_Is}, and for $I_s \ge 3$ $\mu$A there is almost no change in resonance frequency. Choosing the critical current in the range of 5 $\mu$A assuming identical junctions within 2\% would then give a tuning range of $4.96\ge f \ge 3.76$\,GHz.

Another consideration is that $\Phi$ in the expression for $I_s$ is the total magnetic field in the SQUID loop given by $\Phi=\Phi_{ex}-L_{loop}I_{sc}$ where $\Phi_{ex}$ is the external magnetic field, $L_{loop}$ is the inductance of the SQUID loop and $I_{sc}$ is the screening current in the loop \cite{schmidt}. Including the loop inductance and the screening current will also limit the amount that $I_s$ can be suppressed by the magnetic field. Having a small critical current and a small loop inductance reduces this effect.

\begin{figure}
\centering
\includegraphics[width=0.7\textwidth]{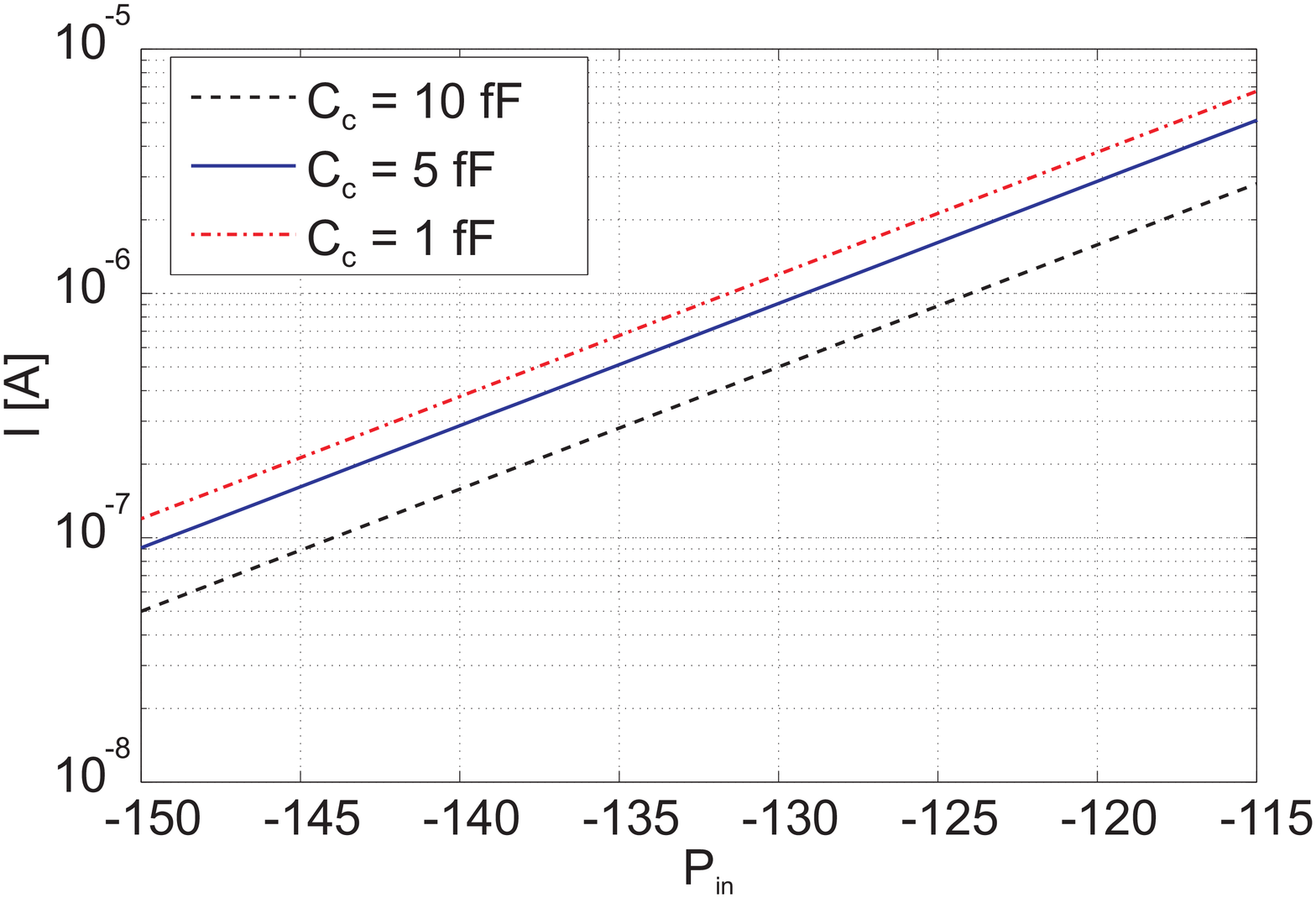}
\caption{\label{fig:I_vs_P} The current at the shorted end of the cavity as a function of applied probe power for different coupling capacitance and no internal losses. The current due to the probe signal should be much less than the minimum value of suppressed critical current of the SQUID in order to be in the linear regime. }
\end{figure}

In order to probe the cavity a signal of a certain power, $P$, has to be applied. The probe signal will cause a current through the SQUID that has to be much less than $I_s$ in order to be in the linear regime. We can derive the expression for the current, at position $x$ along the cavity, as a function of power

\begin{equation}
I(x)=\left(\e^{\gamma x}-\e^{-\gamma x}\Gamma_s\right)\left(\frac{S_{21}\e^{-\gamma \ell}}{1-S_{22}\Gamma_s\e^{-2\gamma \ell}}\right)\sqrt{\frac{8P}{Z_c}}
\end{equation}

where $S_{ij}$ are the scattering matrix for the coupling capacitance and $\Gamma_s$ is the reflection coefficient of the SQUID and $\gamma$ is the propagation constant of the transmission line. Assuming an infinite Josephson energy so that $\Gamma_s=1$ the current at the SQUID end as a function of applied power can be obtained, see figure\,\ref{fig:I_vs_P}. From this we see that in order to be below the minimum of the suppressed current 0.2 $\mu$A the probe power has to be less than -143 dBm for a coupling capacitance of 5 fF.

To increase the tunability an array of $N$ SQUIDs coupled in series could be used for the tuning. The expression for the resonance frequency $f_r$ then modifies to

\begin{equation}
f_r(\Phi)=\frac{f_0}{1+NL_s(\Phi)/L\ell}
\label{eq:f(Phi)}
\end{equation}

where $L_s$ is the inductance of each SQUID. The main advantage of this would be that a higher probe power could be used since $I_s$ of each SQUID can be a factor of $N$ higher at the same detuning \cite{saclay_tune}. Another approach to obtain large tunability is to make the whole center strip into an array of SQUIDs \cite{Lehnert_APL}, the tuning is then archived by tuning the phase velocity of the transmission line. Such a setup however, is not well suitable for qubit coupling since the magnetic field has to be applied to the whole resonator structure and hence also to the qubits during the detuning.

In order to tune the SQUIDs fast, which is necessary for application of quantum gates, a small local field can be used. This is done by current biasing a second transmission line in the vicinity of the SQUIDs. The current $I_B$ needed to apply $0.5\Phi_0$ to the SQUID loop can be approximated by $I_B=\Phi_0 \pi d/A \mu_0$,
where $A$ is the loop area and $d$ is the distance between the bias line and the center of the SQUID loop. The SQUIDs used have an loop area of $\sim$30 $\mu$m$^2$, by placing the bias lines 50 $\mu$ from the SQUIDs  a $I_B \approx 8$ mA is needed (ignoring the effects of flux concentration due to the superconducting ground planes).

\section{Experiments}

To form the Josephson junctions the two layers of Al were deposited from different angles. After the first layer is deposited a small amount of O$_2$ is let in to the deposition chamber ($\sim$1\,mBar) for a short time ($\sim$1\,min) to oxidize the first layer of Al. The oxide forms a few nm thick insulating layer of AlO$_x$ on top of the film. By depositing a second layer of Al after pumping down to base pressure, 5$\times10^{-7}$\,mBar, but from a different angle, the Josephson junctions of the SQUID are formed. The critical current of the Josephson junctions are determined by the area of the junctions, oxidation pressure and the oxidation time. 

\begin{figure}[tb]
\centering
\includegraphics[width=0.39\textwidth]{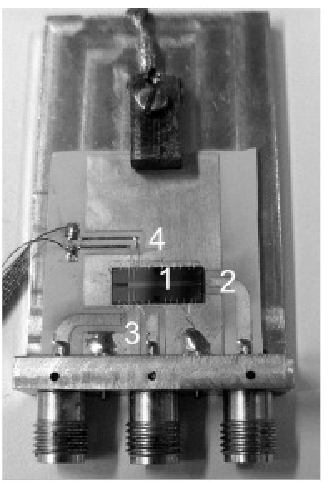}
\includegraphics[width=0.31\textwidth]{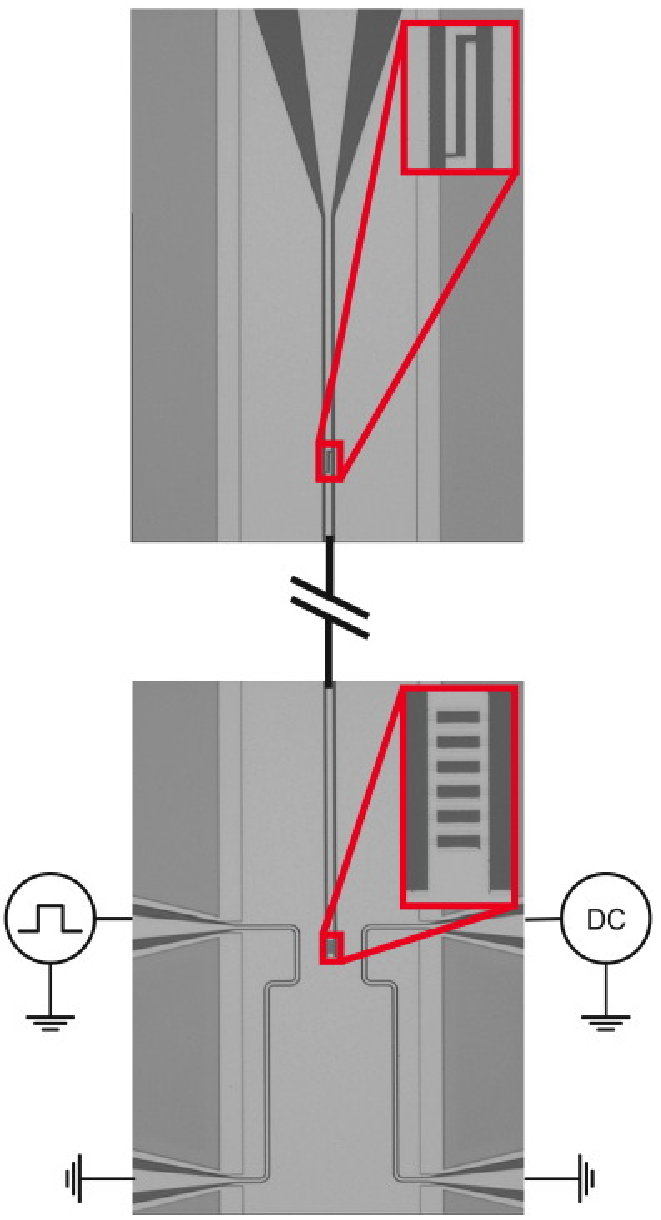}
\caption{\label{fig:sample_hold}(a) The sample holder with a sample mounted. 1 The sample chip. 2 Probe line for the resonator. 3 Lines for fast tuning of the SQUIDs. 4 Lines for DC flux bias of the SQUID.  (b) One sample in close-up with indication of the flux bias. The insets show the array of SQUIDs and the coupling capacitance respectively.}
\end{figure}

The samples were measured in a $^3$He/$^4$He dilution refrigerator with a base temperature below 20\,mK. 
In order to characterize the samples, high frequency semirigid coaxial cables with a characteristic impedance of 50 $\Omega$ were installed in the cryostat. The coaxial cables are stainless steal UT85-SS from the top of the cryostat down to the still level. From the still level down to the mixing chamber  superconducting NbTi UT85 cables are used. Well below its critical temperature the NbTi cables have almost no heat conductivity but a very good electrical conductivity. On each stage the cables are thermally anchored using SMA bulk-head feedthroughs and attenuators to thermalize both the outer and the inner conductors.

\begin{figure}[tb]
\centering
\includegraphics[width=0.5\textwidth]{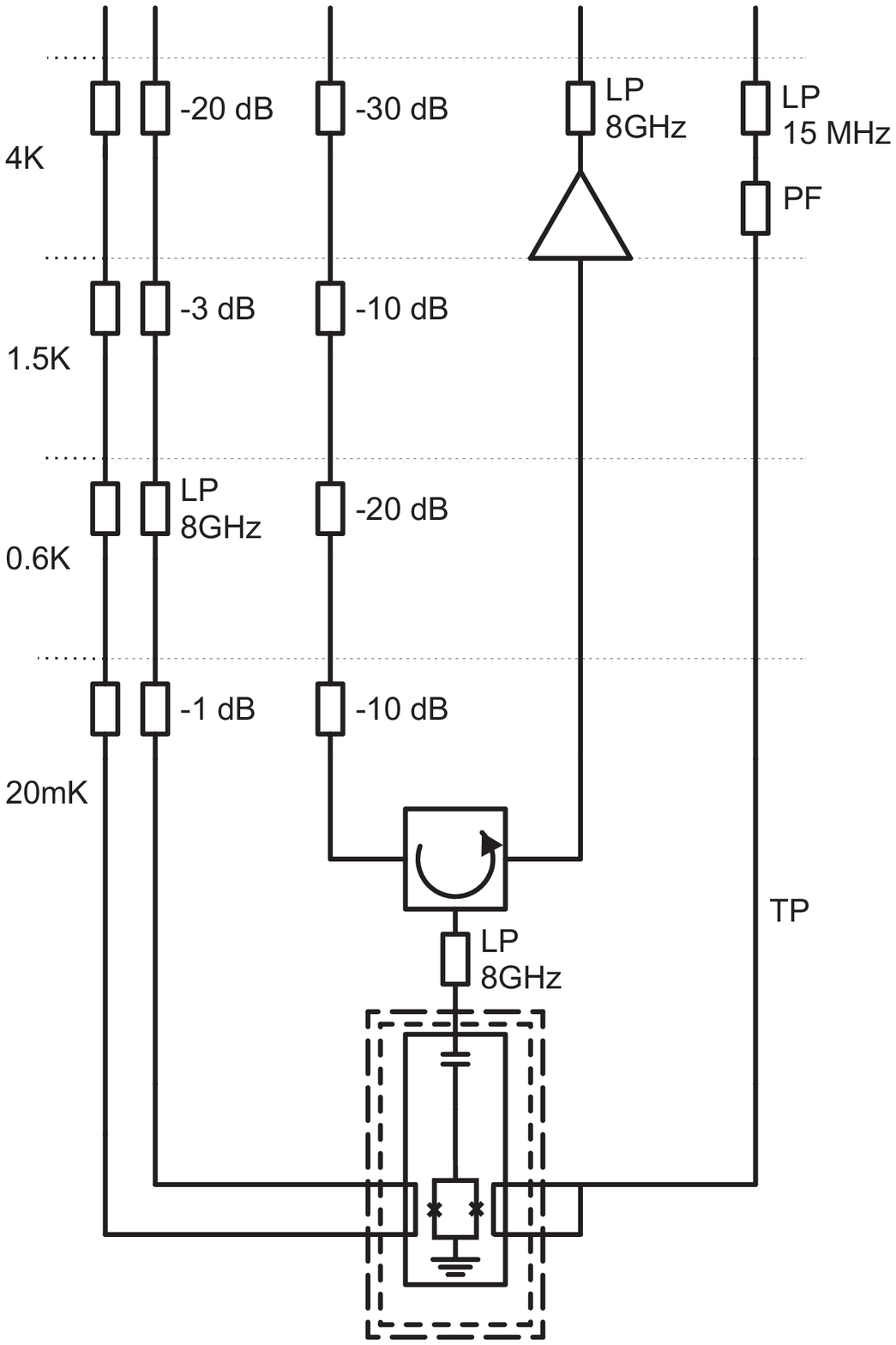}
\caption{\label{fig:mw_setup}Schematic of the microwave setup in the cryostat. The low pass filers (LP) are indicated with their cut of frequency. The rectangles indicates attenuators. The twisted pair (TP) cable, of Nb, used for the DC flux bias is filtered using a 15\,MHz low pass filter and a low pass ($\sim$1\,GHz) powder filter (PF). }
\end{figure}

Two different types of samples were studied, one type with on-chip flux bias and the other one without. The samples with on-chip flux bias was mounted in a sample holder on a PCB circuit board, see figure\,\ref{fig:sample_hold}. On the circuit board, high frequency lines were fabricated to connect the sample, via wire bonds, to the SMA connectors of the sample holder. A twisted pair cable was also used to supply the sample with a DC flux bias. To protect the sample from external magnetic flux noise a magnetic shield consisting of two inner layers of cryoperm and an outer layer of a superconducting Pb was used. For the other type of sample, without an on-chip flux bias, a single SMA connector directly solder to the PCB was used as a sample holder. On the back side of the PCB a coil was then pattern to apply the magnetic field. To minimize heating the coil was coated with a superconducting Sn-based solder. The coil was connected using twisted pair cable.

In order to use low probe power the samples was probed through a circulator with 18 dB of isolation placed at the mixing chamber of the cryostat, see figure\,\ref{fig:mw_setup}. The reflected signal was then amplified by a low-noise cold amplifier mounted to the IVC flange. The amplifier was a Miteq AFS3-04000800-CR-4 with a noise temperature of $\sim$5\,K measured at 4\,K \cite{miteq} . Using this setup the tunability and tuning speed of the devices was studied. The results of these measurements are presented in the next section.

\section{Tunability}
In this following the results obtained from measurements of the fabricated devices are presented. The tunability of four devices and the tuning speed of one tunable transmission lines resonators have been measured as well as the properties of a non-tunable reference device.

\begin{figure}[tb]
\centering
\includegraphics[width=0.4\textwidth]{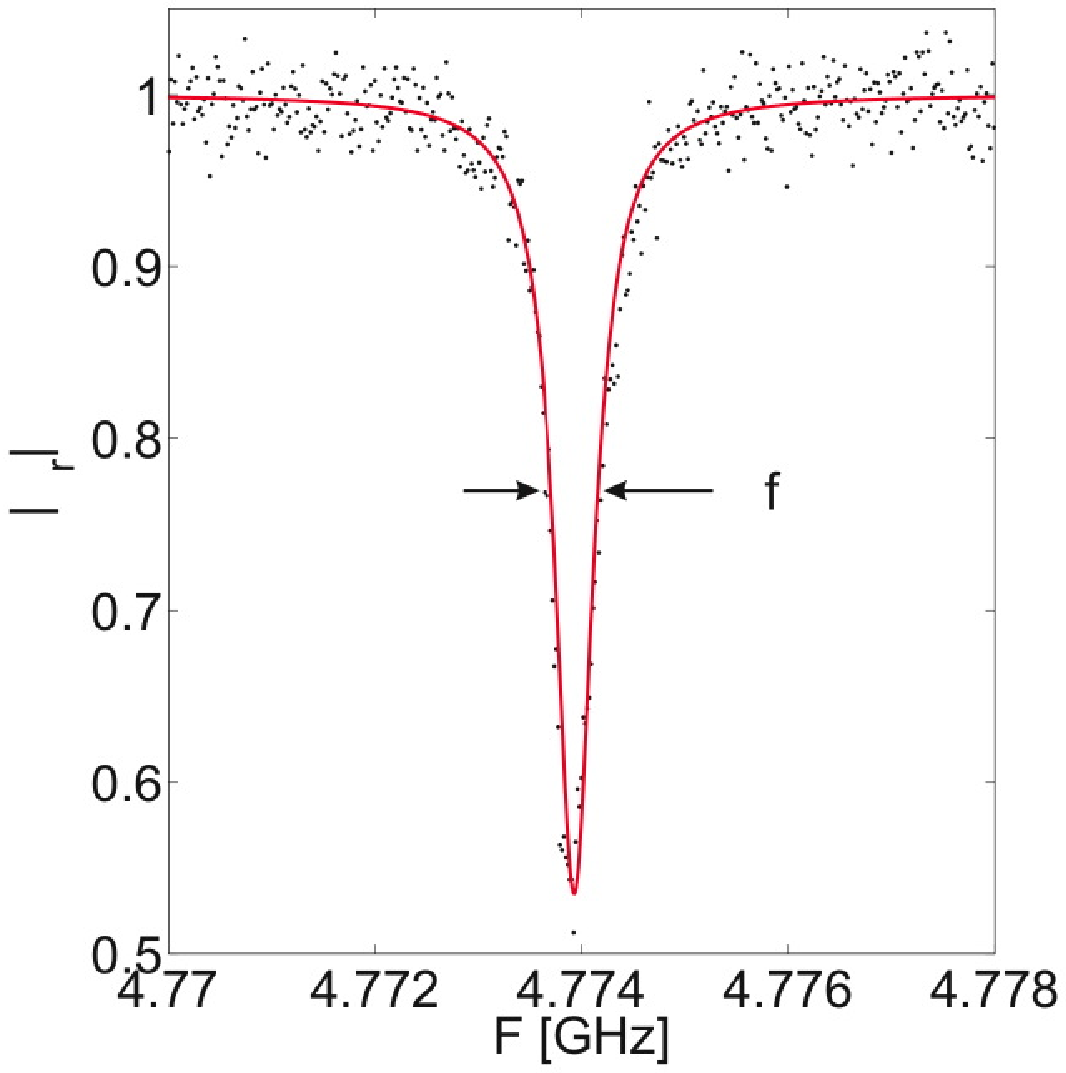}
\includegraphics[width=0.4\textwidth]{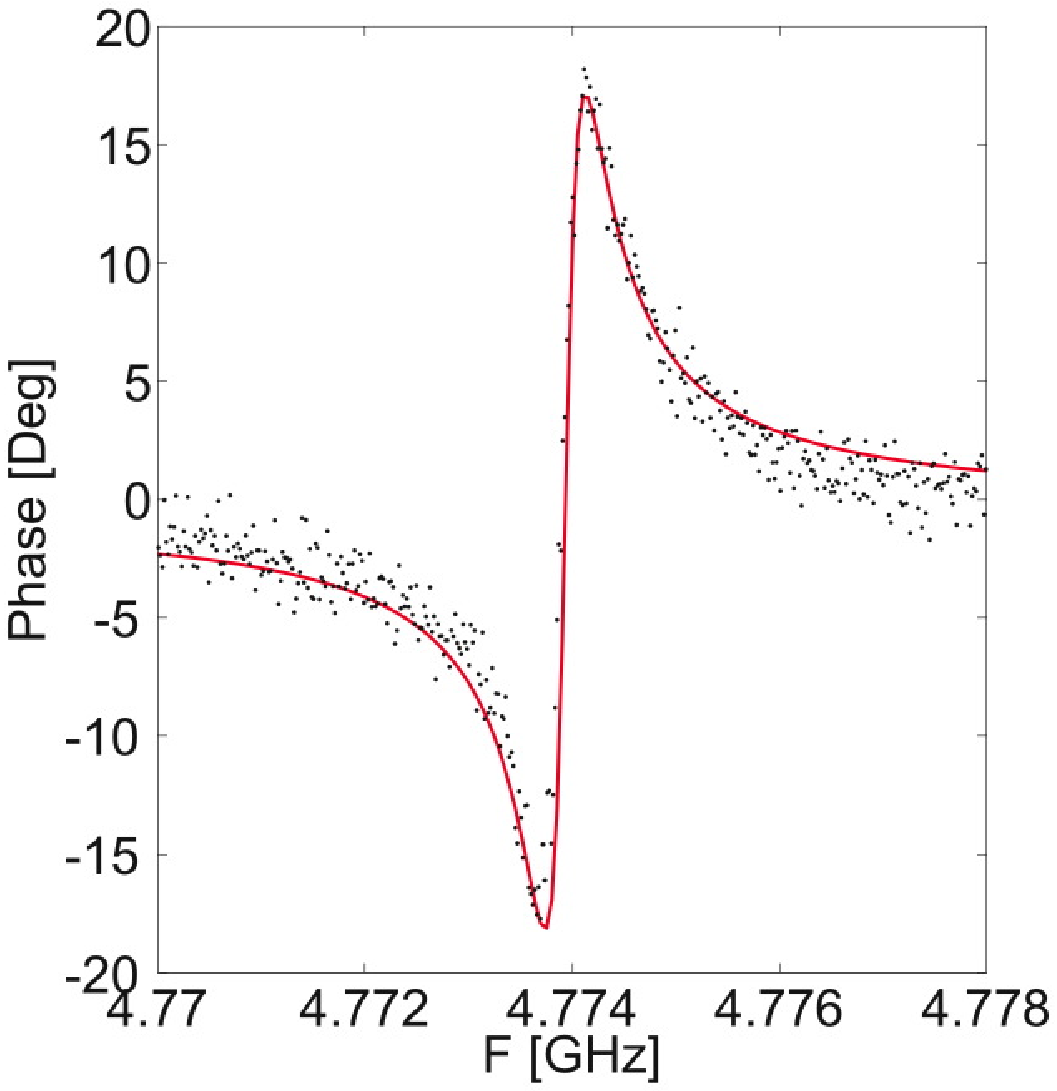}
\caption{\label{fig:resonance}(a) Measured magnitude response for sample C (dots), see table \ref{tab:ABCD}, at zero flux bias using a vector network analyzer. The solid line is a fit of the reflection coefficient $\Gamma_r$. The line width $\delta f$ is the Full Width at half Maximum (FWHM). From the fit a coupling capacitance of 3 fF is obtained. (b) The phase response of the same resonator and the same fit.}
\end{figure}

To measure the scattering parameters of the devices a Vector Network Analyzer (VNA) was used. By sweeping the probe signal over a frequency interval a resonance can be detected either in the phase or in the magnitude of the reflection coefficient. A typical measurement is shown in figure\,\ref{fig:resonance}, where electrical length and losses due to cables and attenuators has been compensated for. From the magnitude response the total $Q$ value can be obtained as $Q_{tot}=f_r/\delta f$, where $\delta f$ is the line width of the resonance and corresponds to the Full Width at Half Maximum (FWHM) of the resonance peak, and $f_r$ is the resonance frequency given by the location of the minimum of the resonance peak. The FWHM and $f_r$ can be obtained by fitting a Lorentzian function to the resonance. The shape of the magnitude and phase response indicated that all the measured resonators were undercoupled, this means that the $Q$ value is dominated by the internal losses and not by the coupling, \textit{i.e.} $Q_{int}<Q_{ext}$. The total reflection coefficient $\Gamma_r$ of the resonator is written as

\begin{equation}
\label{eq:Gamma_r}
\Gamma_r=S_{11}+\frac{S_{12}S_{21}\Gamma_se^{-\gamma \ell}}{{1-S_{22}\Gamma_se^{-2\gamma \ell}}}
\end{equation}

By fitting $\Gamma_r$ to the measurements, assuming only dielectric losses, the parameters of the resonator can be inferred. In order to obtain a god fit, a coupling capacitance of 3\,fF for 50\,$\mu$m long fingers and a capacitance of 154\,pF/m has to be assumed, slightly less than the design value 164\,pF/m. This gives an external 
$Q$ value of $Q_{ext}=65000$.

\begin{figure}[tb]
\centering
\includegraphics[width=0.385\textwidth]{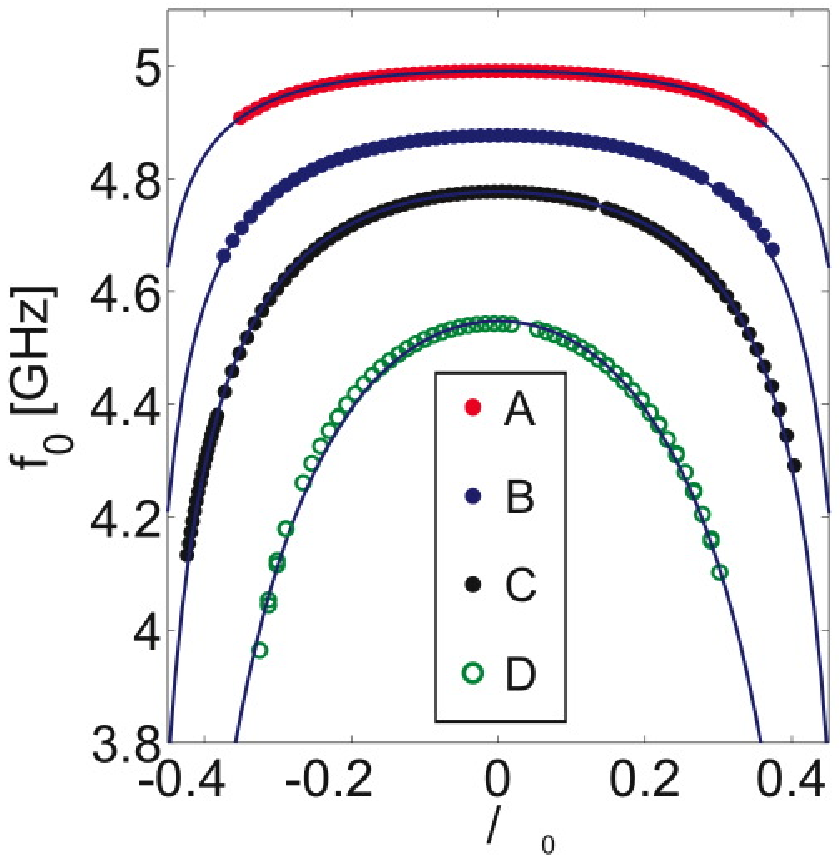}
\includegraphics[width=0.415\textwidth]{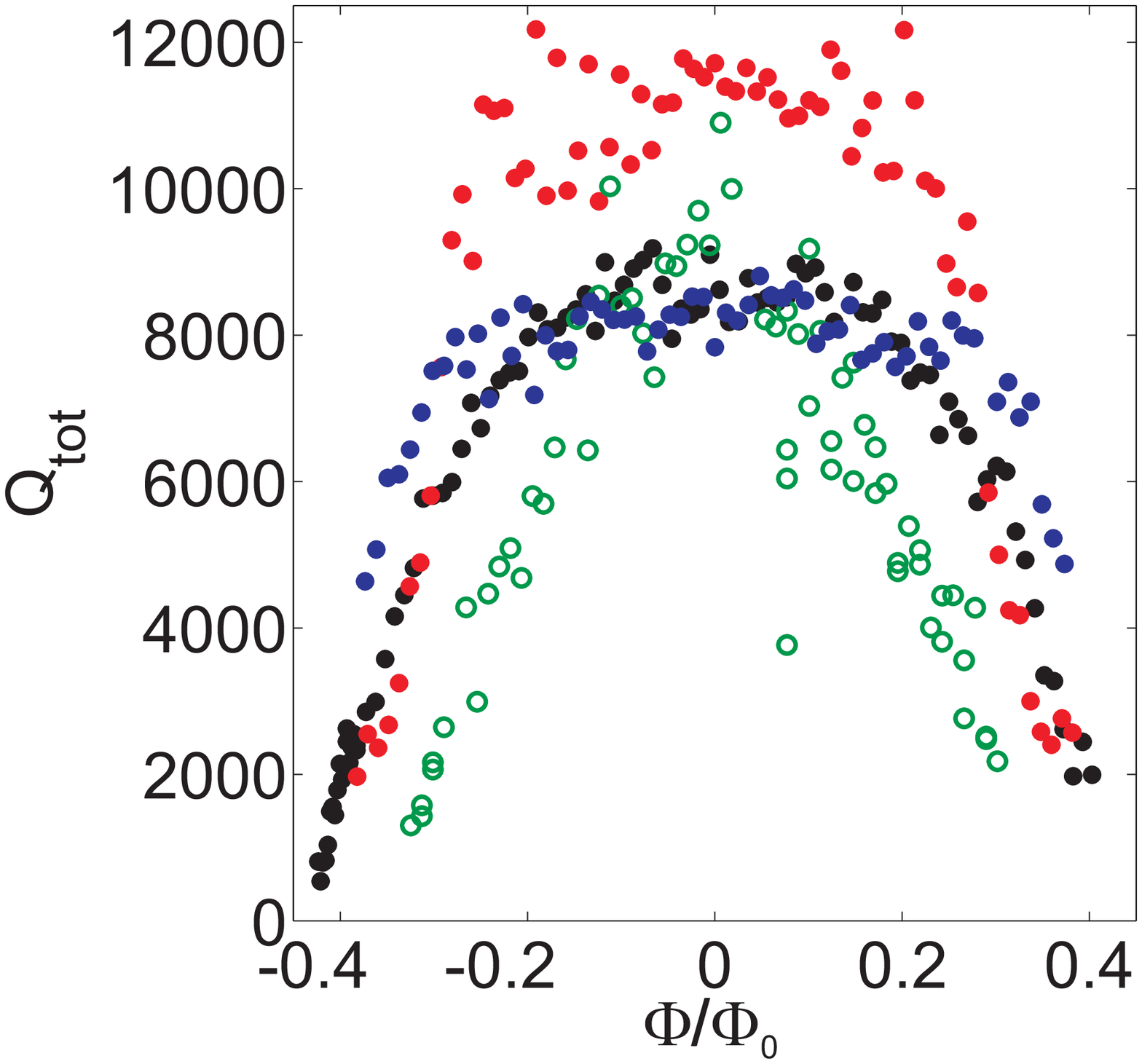}
\caption{\label{fig:tune}
(a) The measured resonance frequency together with a fitted curve as a function of applied magnetic flux for the samples A, B, C and D. The parameters of the samples are summarized in table \ref{tab:ABCD}. (b) The Q value measured from the line width of the resonance curve for sample A, B, C and D as a function of applied magnetic flux $\Phi$. The $Q$ value decreases rapidly when the applied flux approaches  0.3$\Phi_0$ for all four samples. }
\end{figure}

\begin{table}[bt]
\centering

\begin{tabular}{|c|c|c|c|c|c|c|}
  \hline
  Sample&  $I_c$ &  $N$ &   $\ell_c$    & $f_0$   & $L_s(0)/L\ell$ & $\Delta f$ \\ \hline
                & $\mu$ &          & $\mu$m & GHz     &                        & MHz         \\ \hline
  A & 4     &  1       &  25     &   5.062   & 0.0141 & 91 \\ \hline
  B & 2     &  1       &  50    &   5.025    & 0.0303 & 265 \\ \hline
  C & 1.2   &  1     &  50    &   5.034    & 0.0537 & 744 \\ \hline
  D & 2.4   &  6     &  50    &   5.350    & 0.0292 & 580  \\ \hline
\end{tabular}
\caption{\label{tab:ABCD} Parameters for samples A, B, C and D. $I_c$ is the critical current of each junction in the SQUID, $N$ is the number of SQUIDs in series,  $\ell_c$ is the length of the coupling capacitance fingers, $f_0$ is the $\lambda/4$ resonance frequency, $L_s(0)$ is the zero flux SQUID inductance, $L$ is the inductance per unit length of the transmission line and $\ell$ is the length if the resonator.}
\end{table}

To measure the tunability of the devices a DC magnetic flux bias was applied to the SQUID. In figure\,\ref{fig:tune}(a) the resonance frequency as a function of applied magnetic flux for four samples A, B, C and D are shown. The sample parameters are summarized in table \ref{tab:ABCD}. The expression for the resonance frequency as a function of flux is given by eq.\,\ref{eq:f(Phi)}
which is fitted to the measured data with a good accuracy and shown as solid lines in figure\,\ref{fig:resonance}. 
In figure\,\ref{fig:tune}(b) we have plotted the extracted Q-value as a function of flux. As can be seen the Q-values are typically of the order of $10^4$ at zero flux and it decreases for increasing flux.

\begin{figure}
\centering
 \includegraphics[width=0.6\textwidth]{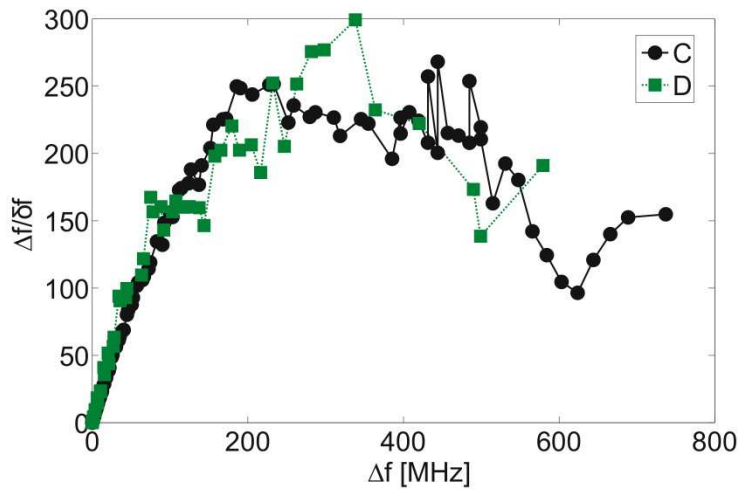}
\caption{\label{fig:delta_f} The number of line widths $\delta f$ that the devices can be detuned as a function of the detuning $\Delta f$ for sample C and D. For both samples a maximum detuned line widths obtained is around 250.}
\end{figure}

As a figure of merit the tunability can be measured in the number of line widths that the device can be detuned. Since the line width increases with the detuning this has to be compensated for. In figure\,\ref{fig:delta_f} the number of line widths detuned as a function of detuning is shown for sample C and D. The highest value obtained is around 250 for both samples.
The tunability observed for the samples is less than what could be expected if the tunability was only limited by the asymmetry of the SQUID junctions. The limiting factor of the tunability is the strong decrease in $Q$ value as the applied flux approaches $\Phi = 0.3\Phi_0$. 

\section{Reference device}
\begin{figure}
\centering
 \includegraphics[width=0.4\textwidth]{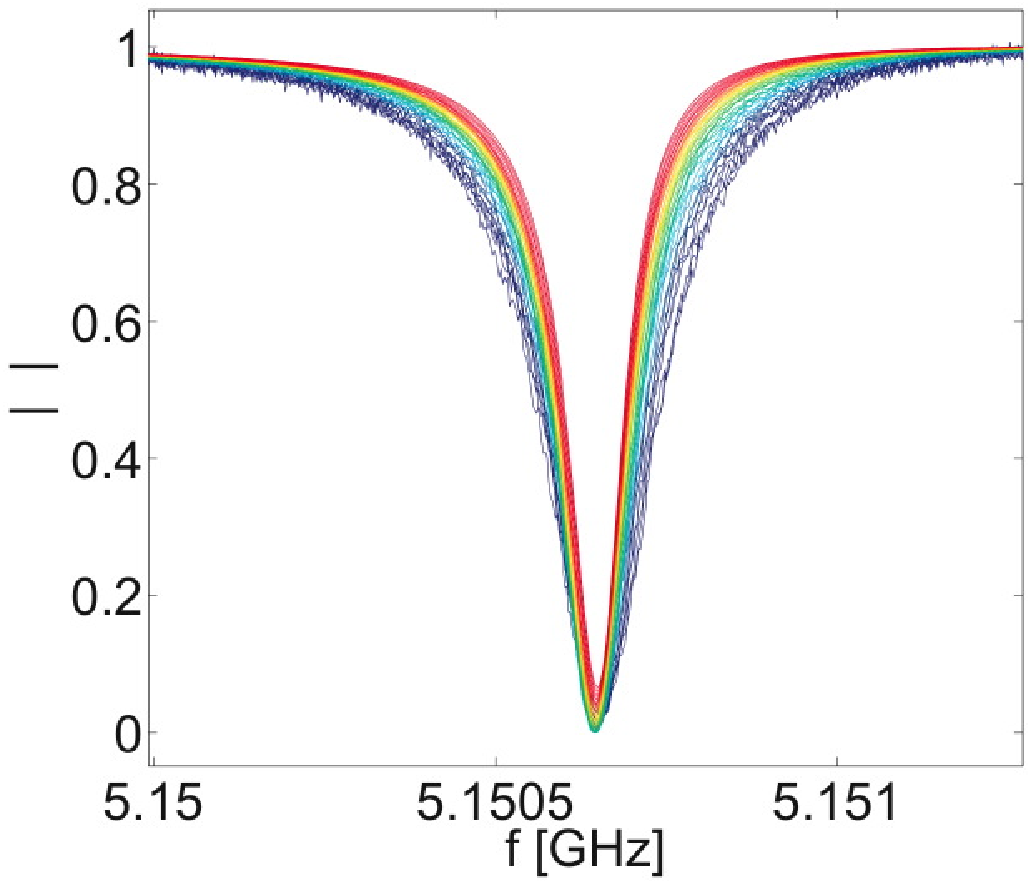}
 \includegraphics[width=0.4\textwidth]{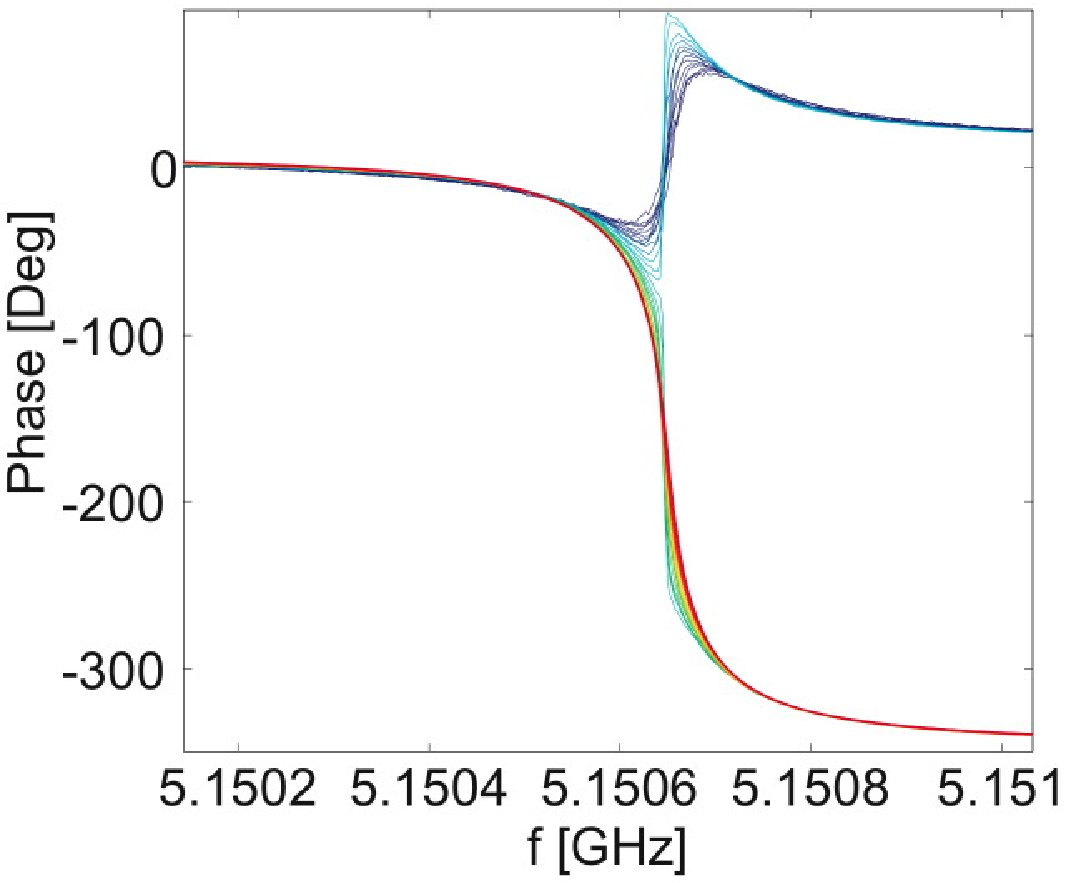}
\caption{\label{fig:ref}
(a) The magnitude response of the reference device as the power is increased. Blue indicates low power and red high power. (b) Phase response of the reference device as the power is increased. The shape of the phase response goes from an under coupled shape (blue) to an over coupled shape (red). }
\end{figure}

\begin{figure}[b]
\centering
\includegraphics[width=0.4\textwidth]{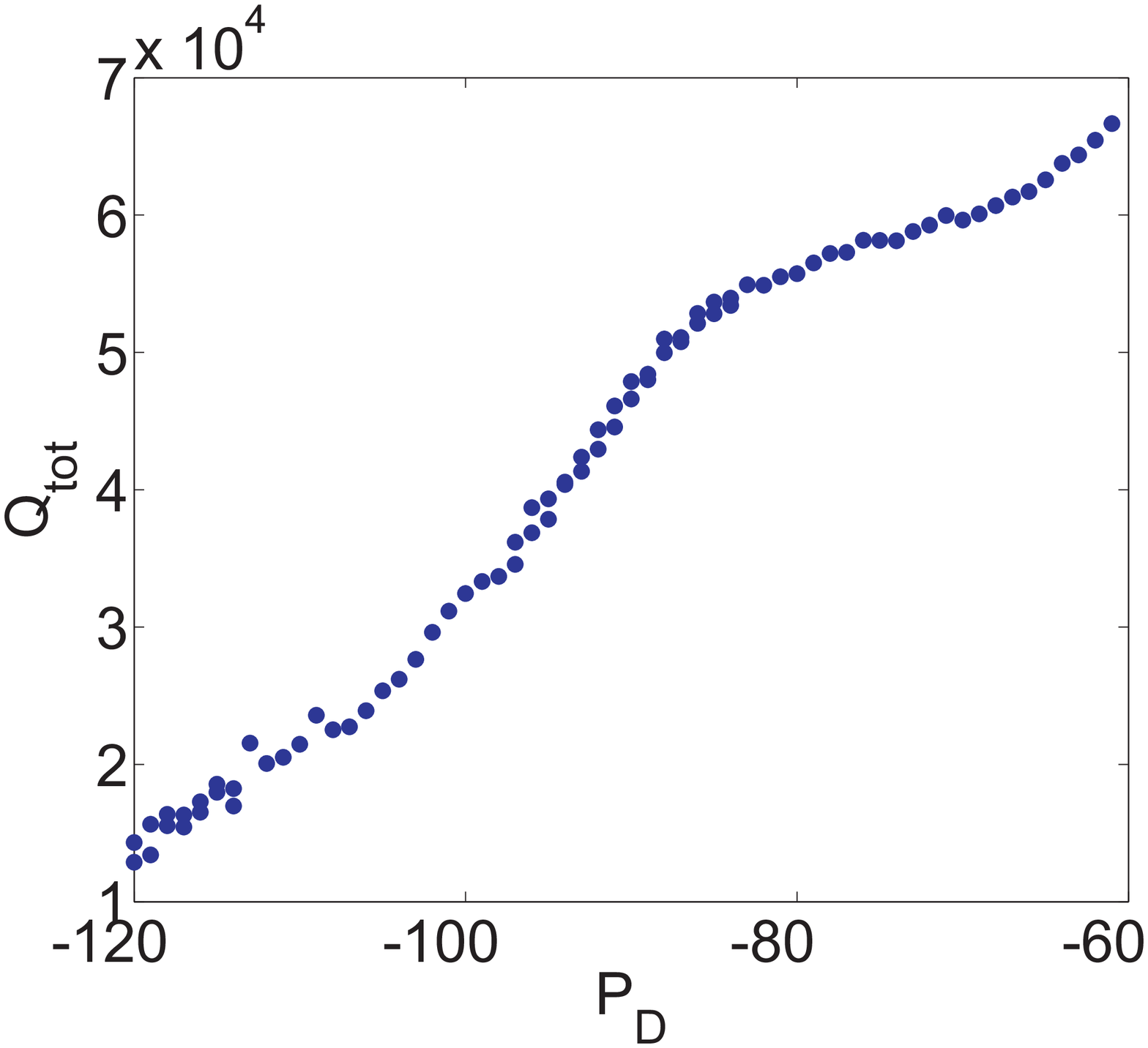}
\includegraphics[width=0.4\textwidth]{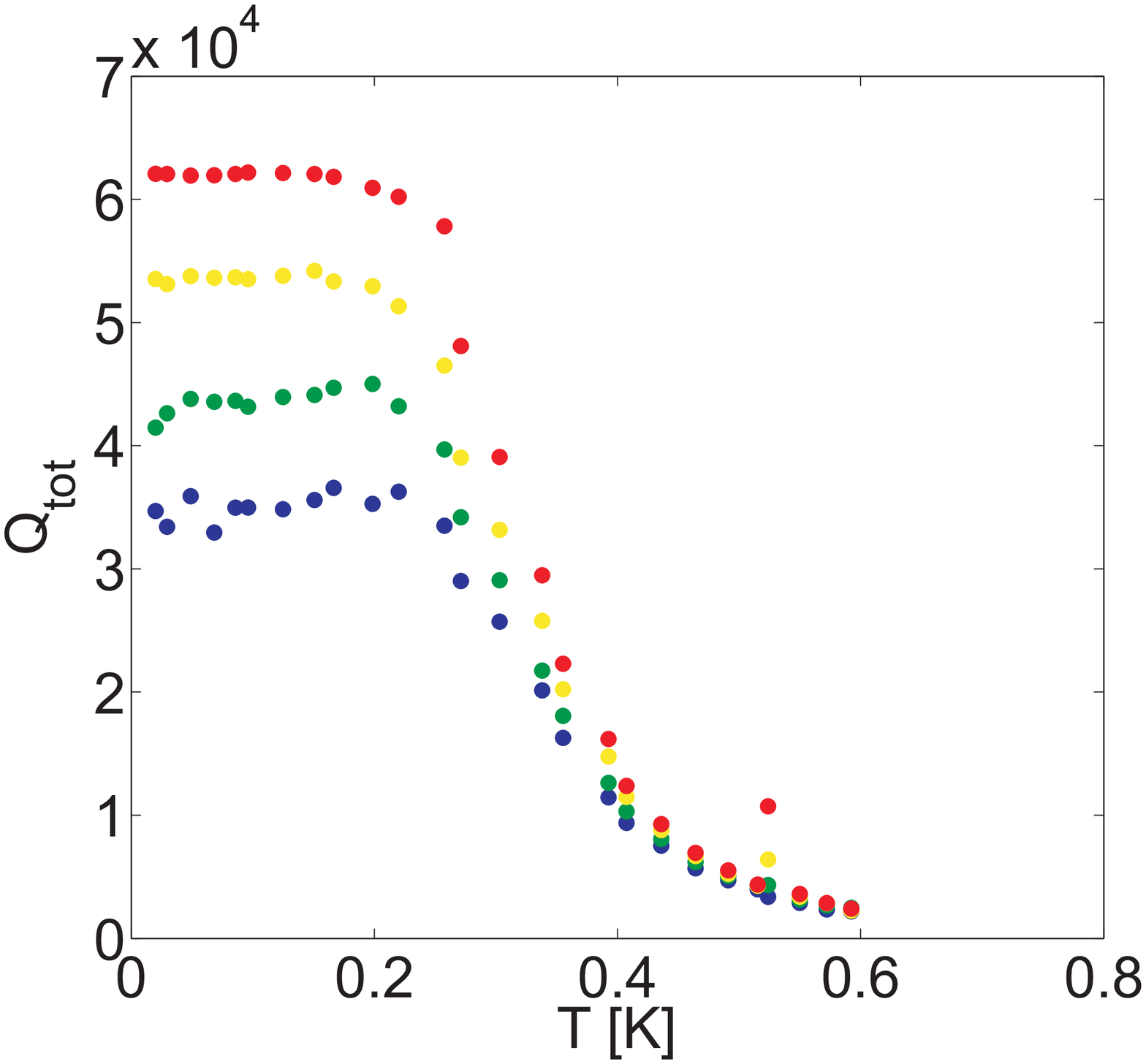}
\caption{\label{fig:Q}
(a) Q value as a function of applied power. As the power is increased the Q value increases. This power dependence is consistent with similar experiments where the power dependence were contributed to dielectric losses. (b) The  temperature dependence of the Q value measured at different drive power. At temperatures above $\sim$0.2 K the Q value starts to be dominated by resistive losses in the aluminum film.}
\end{figure}

To better understand the source of dissipation inside the resonator a reference device without any SQUID was fabricated. The reference device was measured in the same setup as the other samples. When measuring the response at low drive powers this device was also found to be undercoupled. As the measurement power was increased a transition from undercoupled to overcoupled response was  observed, see figure\,\ref{fig:ref}. In the undercoupled case the phase goes down a little bit and then back up again while for the overcoupled case the phase goes from 0$^\circ$ to -360$^\circ$ on resonance. By fitting the reflection coefficient, eq.\,\ref{eq:Gamma_r}, setting $\Gamma_s = -1$ for a short circuit, a coupling capacitance of about 2 fF was obtained for 25 $\mu$m fingers. This gives a $Q_{ext} = 1.5\times10^5$ . In figure\,\ref{fig:Q}(a) it is seen that the Q value increases with the applied power. The resonator goes from $Q = 10^4$ to $Q =7\times10^4$ as the power is increased from -120 dBm to -60 dBm. Due to limitations in the VNA used, more power was not possible to apply without first warming up the cryostat to remove cold attenuators, this was however not done. An increase in $Q$ as a function of increased drive power was also measured by Martinis  \textit{et al.} \cite{Martinis_PRL_2005} for a lumped element resonator. They attributed this effect to the existence of two-level systems in the dielectric of their capacitor. As the power is increased the two level systems start to be saturated. As more and more two-level systems get saturated less energy can be absorbed from the resonator and the $Q$ increases. The amount of two-level systems present in the dielectric depends on the material used and the quality of it. In order to improve the $Q$ value at low drive power a different dielectric material should be considered.

The temperature dependence of the $Q$ value was also investigated. For low temperatures, T $\le$ 0.2 K the Q value is independent of temperature, see figure\,\ref{fig:Q}(b). As the temperature is increased further the Q value starts to decrease in a way that is consistent with an increase in resistive losses due to thermally excited quasi-particles.
Even though the loss mechanisms in our cavities are not fully understood we can say the following. For the cavities presented in this paper our Q-values are limited by intrinsic losses. The thermally excited quasi particles do not seem to limit our Q-value below 200\,mK. 
For zero flux it seems as if the Q-value seems to be due to dielectric losses, but at increased flux there is a different mechanism. 
The flux noise needed to explain the decrease in Q at large flux would have to bee quite large. Considering our magnetic shielding we do not think it is likely that flux noise is the problem. Instead we believe that the reduction of Q for higher flux is due to the sub-gap resistance in the SQUID junctions. Assuming a sub-gap resistance of approximately $800\,\Omega$ would explain the reduction in Q. We would like to point out that the SQUID junctions have relatively high critical current densities which would also lead to a relatively low sub-gap resistance.

\section{Fast tuning}

\begin{figure}
\centering
\includegraphics[width=0.7\textwidth]{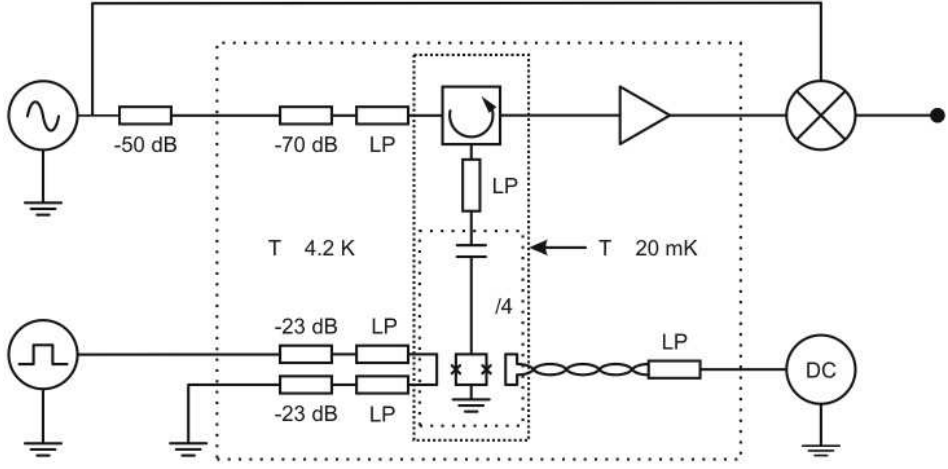}
\caption{\label{fig:setup} 
Measurement set up for the tuning speed experiment. First the cavity is excited on resonance at one flux bias. When the cavity is filled with energy it is detuned rapidly, by a flux pulse, to a new resonance frequency several line widths from the drive. When the resonator is detuned from the drive no more energy is put in to the resonator and the energy inside the cavity starts to leak out through the coupling capacitance. The leakage of energy creates a signal that can be detected using a mixer and a fast oscilloscope. }
\end{figure}

To do fast quantum gates one has to be able to tune the device fast. A measurement scheme where the reflection coefficient is probed as the resonator is detuned is limited by the ring up time of the resonator  $Q/\omega_r$. For a 5\,GHz resonator with a $Q = 10^4$ the ring up time is $\sim$300 ns.  In order to measure tuning speeds faster than this a new measurement setup was implemented. In the new set up, see figure\,\ref{fig:setup}, the drive signal is split up. One part of the signal goes down into the cryostat through the circulator and down to the sample. The signal from the resonator goes up through the circulator, through the cold amplifier and in to a mixer at room temperature. The signal from the resonator is there mixed with the other part of the drive signal. The output from the mixer is then filtered through a low pass filter and recorded on a fast oscilloscope. The measurements were performed using a drive power as low as -145 dBm giving approximately 1 photon on average in the cavity. This low photon number was possible using several million averages.

\begin{figure}[tb]
\centering
 \includegraphics[width=0.4\textwidth]{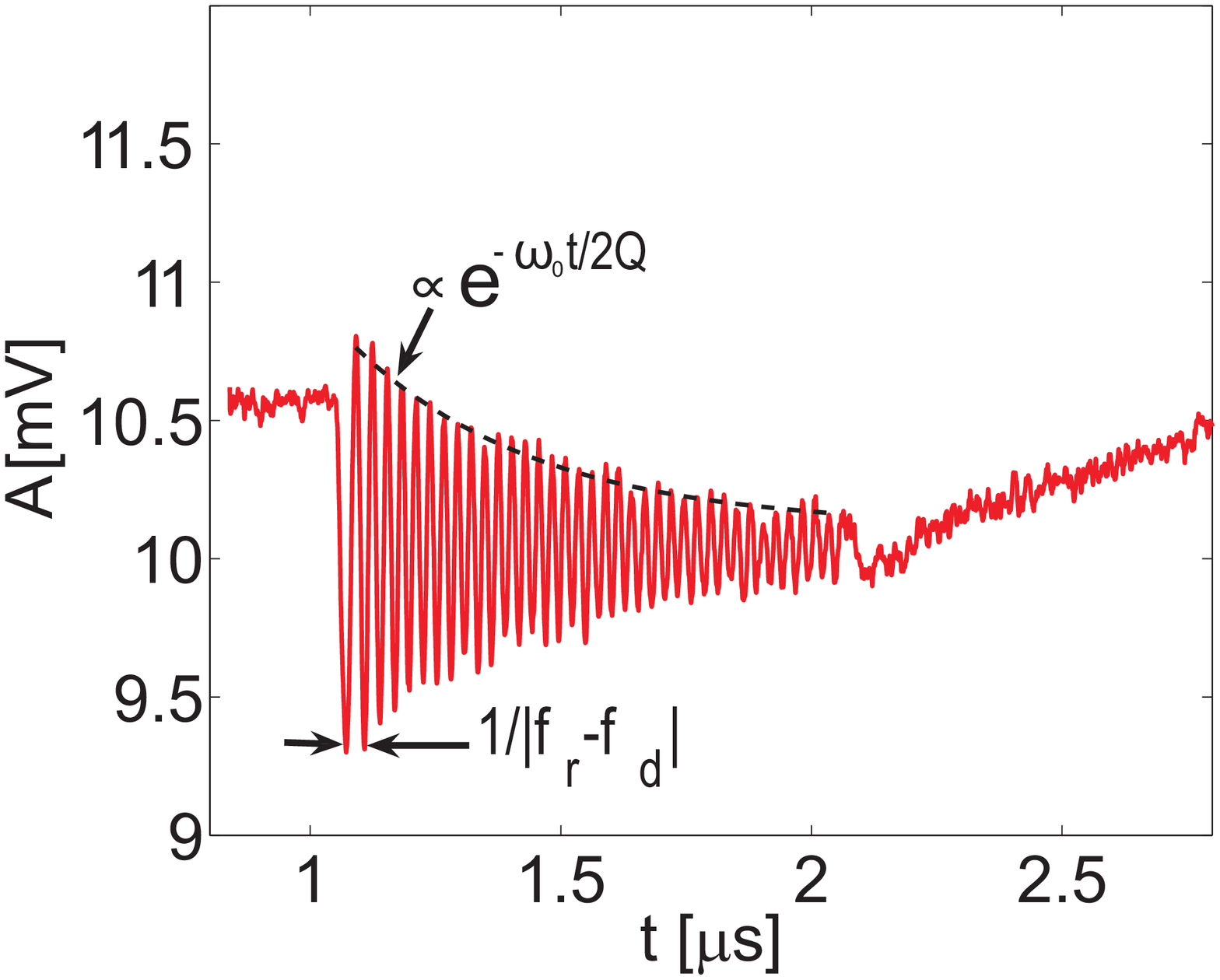}
 \includegraphics[width=0.4\textwidth]{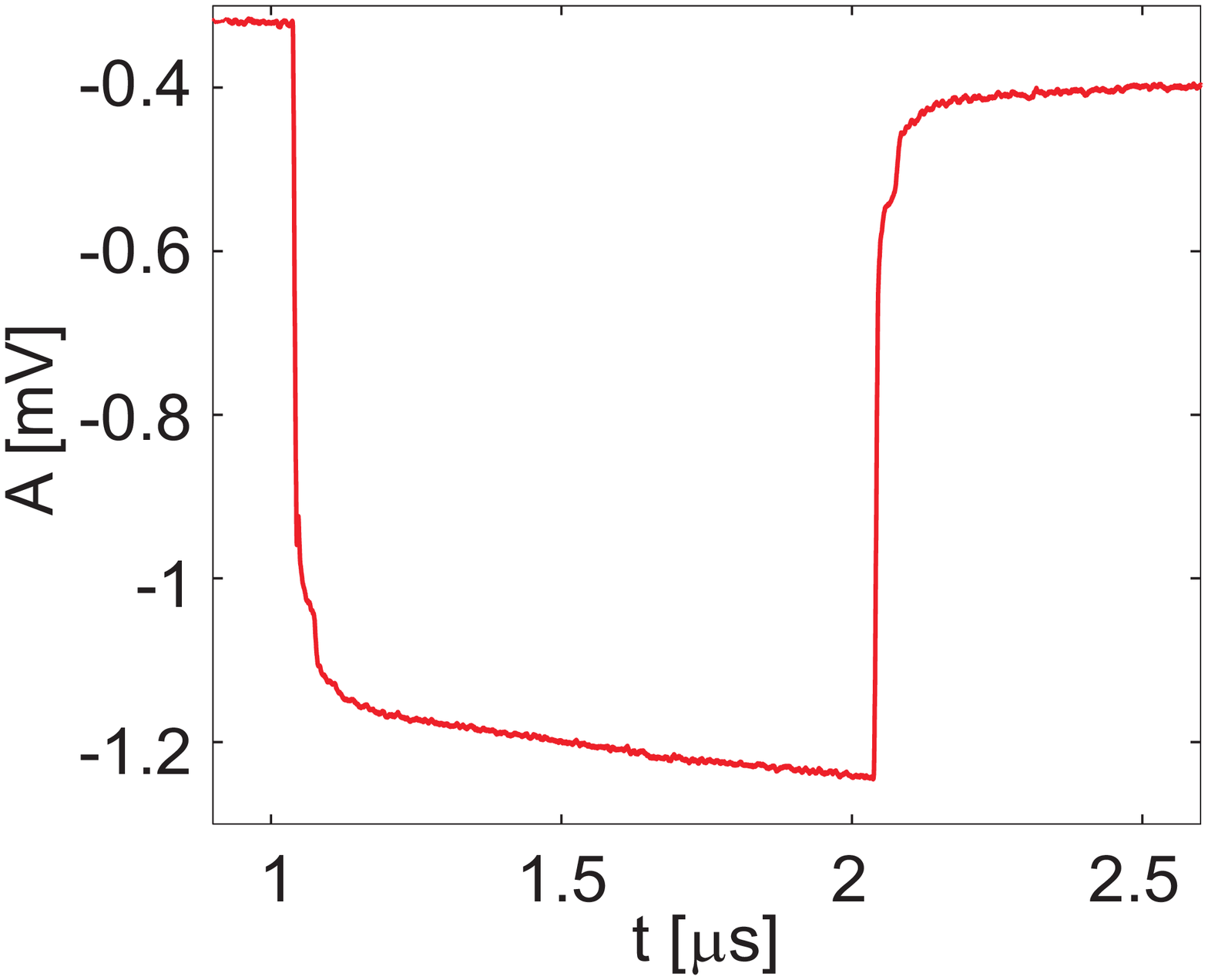}
\caption{\label{fig:stretch}
(a) A measurement of resonator using the fast tuning set up in figure\,\ref{fig:setup}. From these measurements the detuning $|\omega_n-\omega_d|$ and the $Q$ value of the device can be measured. (b) The applied fast pulse used to detune the resonator.}
\end{figure}

The resonator is tuned, using the DC flux line, in to resonance with the drive signal $V_d\sin\omega_dt$. On resonance the resonator builds up energy until a steady state is reached. A fast rectangular flux pulse is then applied to the resonators fast flux line that shifts the resonance frequency to a new frequency $\omega_n$. The energy stored in the resonator has to adjust its frequency to $\omega_n$ in order to match the new resonance condition. If $|\omega_d-\omega_n|\gg \omega_d/Q$ no more energy is put into the resonator by the drive signal. The energy that is stored inside the resonator before the detuning starts to leak out through the coupling capacitance. The leakage causes a signal from the resonator, $V_n(t)\sin \omega_n t$. The signal that leaks out is amplified by the cold amplifier and then put in to the mixer at room temperature. In the mixer the signal from the resonator is multiplied with the other part of the drive signal giving the output signal

\begin{equation}
V_{mix}(t) \propto V_n(t)\cos(\omega_n-\omega_d)t+V_n(t)\cos(\omega_n+\omega_d)t.
\end{equation}

The high frequency component $|\omega_n+\omega_d|$ of the signal is filtered out using a low pass filter , and the remaining signal can be measured using a fast oscilloscope. The time dependence of the amplitude $V_n(t)$ is governed by the decay of the energy stored in the resonator

\begin{equation}
V_n(t) \propto e^{\left(-\frac{\omega_nt}{2Q}\right)}
\end{equation}

where the factor of 2 comes from the fact that amplitude and not power is measured. From these measurements, the detuning $|\omega_n-\omega_d|$ and the $Q$ value can be obtained, see figure\,\ref{fig:stretch}.

\begin{figure}
\centering
 \includegraphics[width=0.385\textwidth]{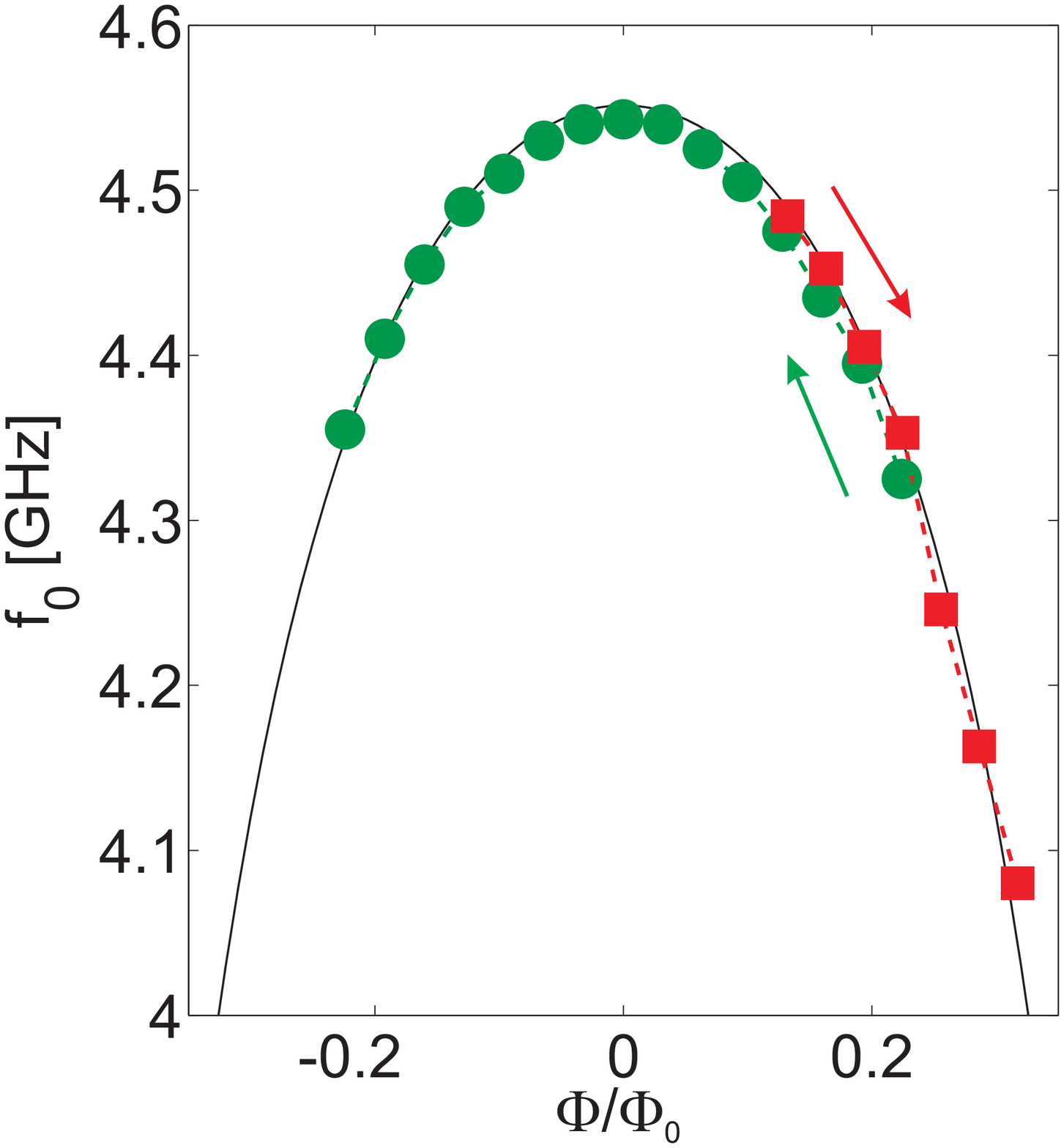}
 \includegraphics[width=0.415\textwidth]{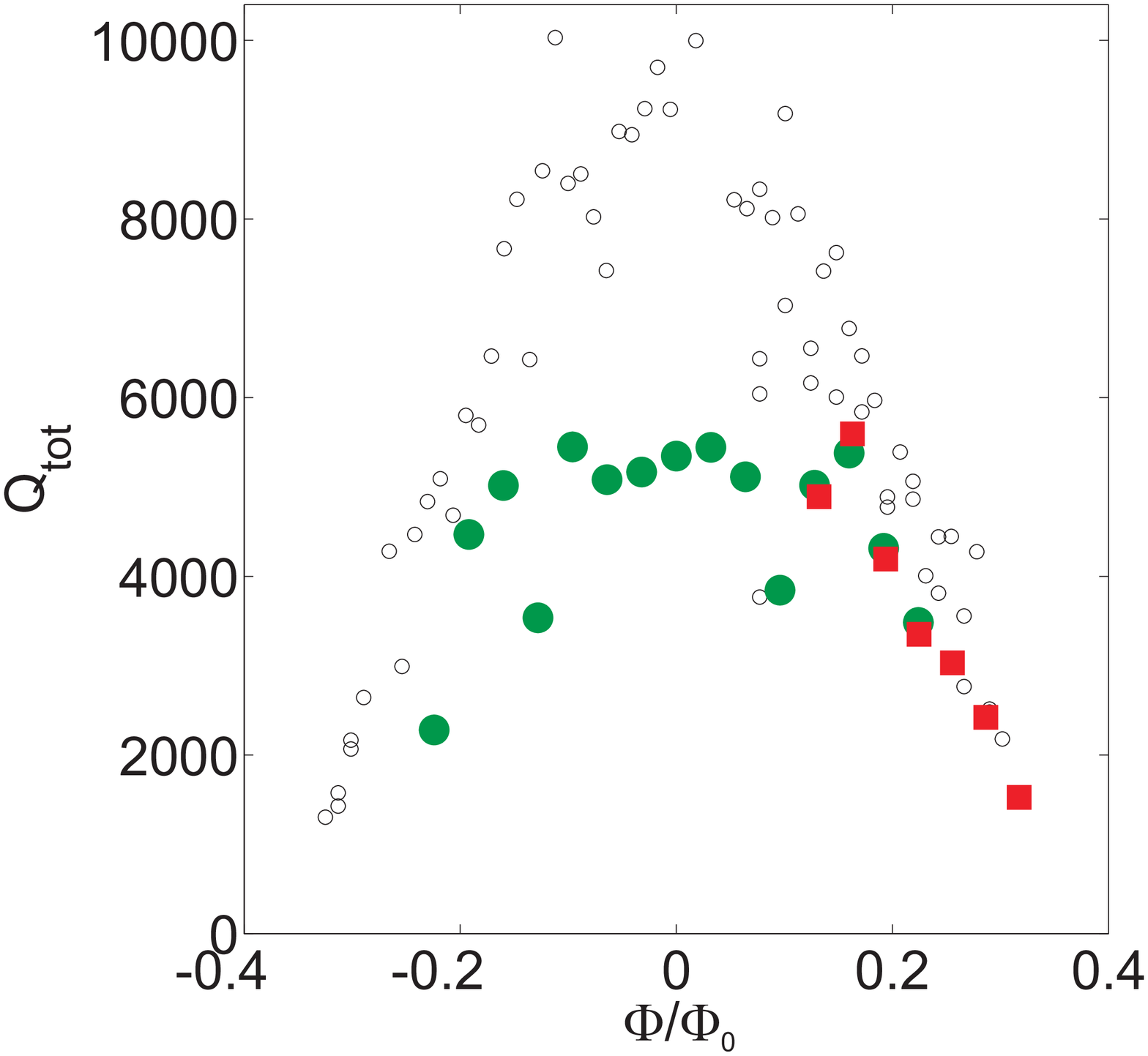}
\caption{\label{fig:fast}
(a) The resonance frequency as a function of flux using fast flux pulses. The solid line is the calculated resonance frequency. Both negative (circles) and positive (squares) flux pulses can be applied. The resonance frequency can hence both be increased and decreased. (b) The obtained $Q$ value from the line width measurements (open circles) and from the decay measurement (solid).}
\end{figure}

By varying the flux pulses amplitude the resonance frequency as a function of flux can be mapped out, see figure\,\ref{fig:fast}. Positive and negative pulses can be applied which means that the frequency can be shifted both to higher  and lower values. In order to increase the frequency, work has to be done on the cavity by the applied field while in the case of a decrease in frequency the cavity does work on the field used to tune the resonator.
The $Q$ value as a function of flux pulse is showed in \ref{fig:fast} together with the data obtained from measurements of the line width. Due to poor signal to noise ratio the measurement had to be performed several times and averaged, as mention previously.

The $Q$ value obtained from the decay time measurements differ substantially from the $Q$ values obtained from the line width around zero flux bias. To understand this discrepancy a histogram of the pulse amplitude from the pulse generator was measured. Using this histogram a probability distribution for the amplitude height can be calculated. Since the amplitude of the flux pulse determine the frequency $\omega_n$ a distribution in pulse amplitude can be converted into a distribution in frequency. The average measured signal $S(t)$ is obtained as

\begin{equation}
\label{eq:S}
S(t)\approx e^{-\omega_n t/2Q}\sum_k^N P_k\sin(\omega_k-\omega_d)t
\end{equation}

where $N$ is the number of bins used in the histograms $P_k$ is the probability of obtaining the frequency $\omega_k$. Using the measured histogram the signal $S(t)$ can be calculated. If a function $e^{-\omega_n t/(2Q)}$ is fitted to $S(t)$ a $Q \approx 5000$ is obtained even if a $Q = 10000$ was used in the calculations \ref{eq:S}. This calculations suggests that the imperfections of the pulse generator used to create the flux pulses is the cause of the degradation seen in the measured $Q$ value.

\begin{figure}[t!]
\centering
 \includegraphics[width=0.4\textwidth]{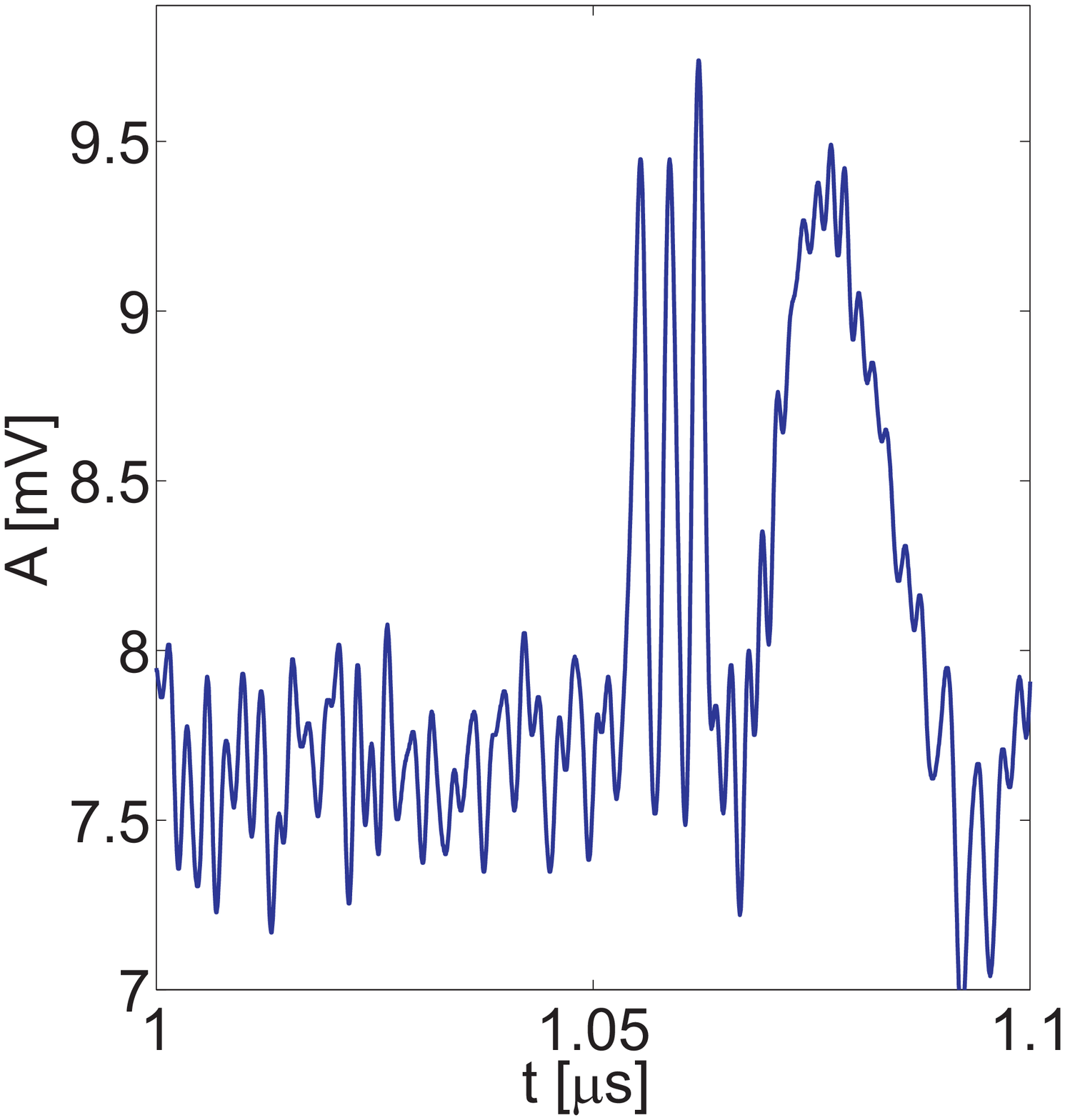}
 \includegraphics[width=0.4\textwidth]{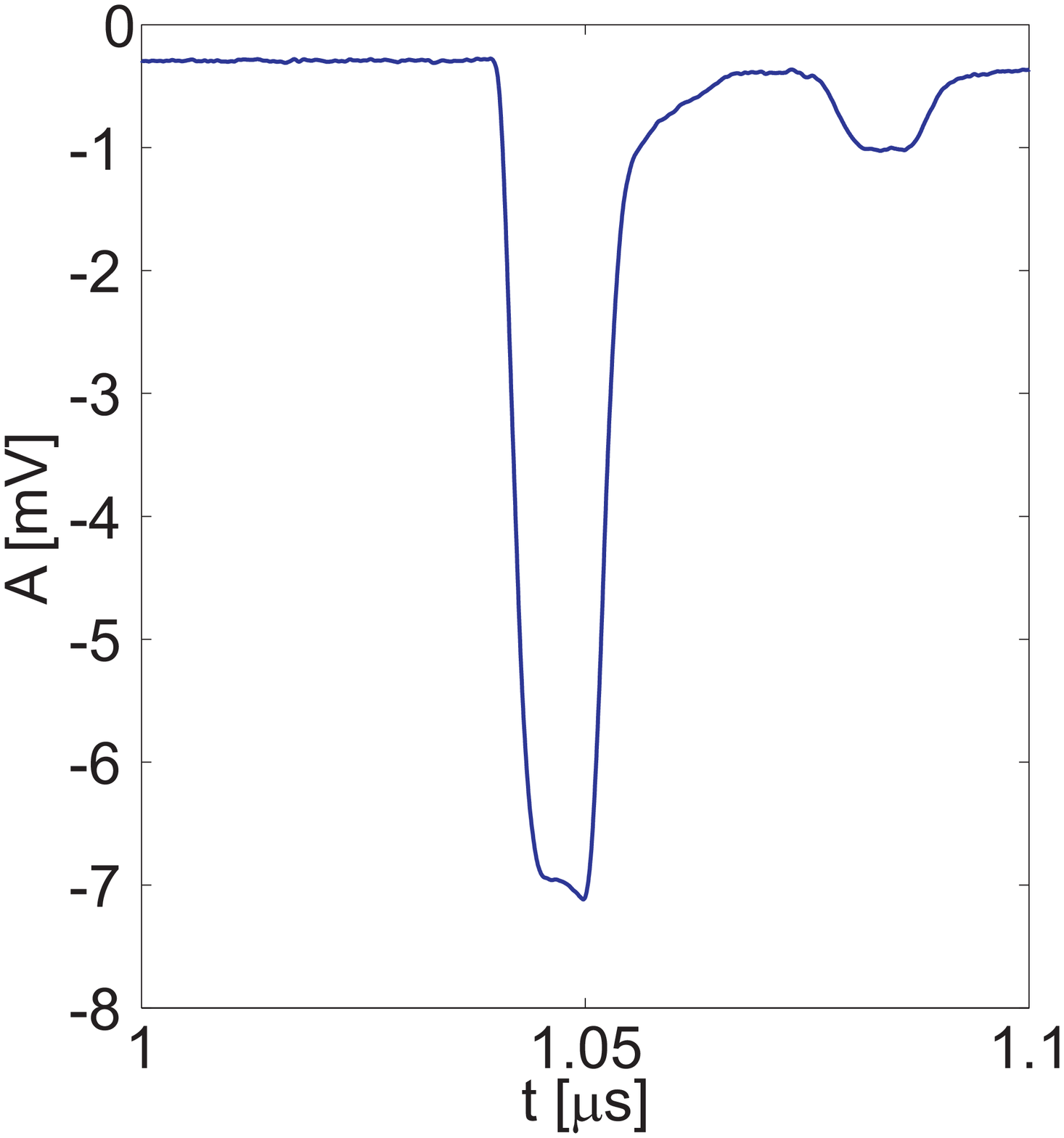}
 \caption{\label{fig:10ns}
(a) Trace obtained for a 10 ns long flux pulse. Three peaks of the oscillations observed, the distance between the peaks gives a detuning of 330\,MHz. After the oscillations there some slower structure due to imperfections in the applied pulse. (b) The 10 ns applied pulse measured with a fast oscilloscope.}
\end{figure}

\begin{figure}[t!]
\centering
 \includegraphics[width=0.4\textwidth]{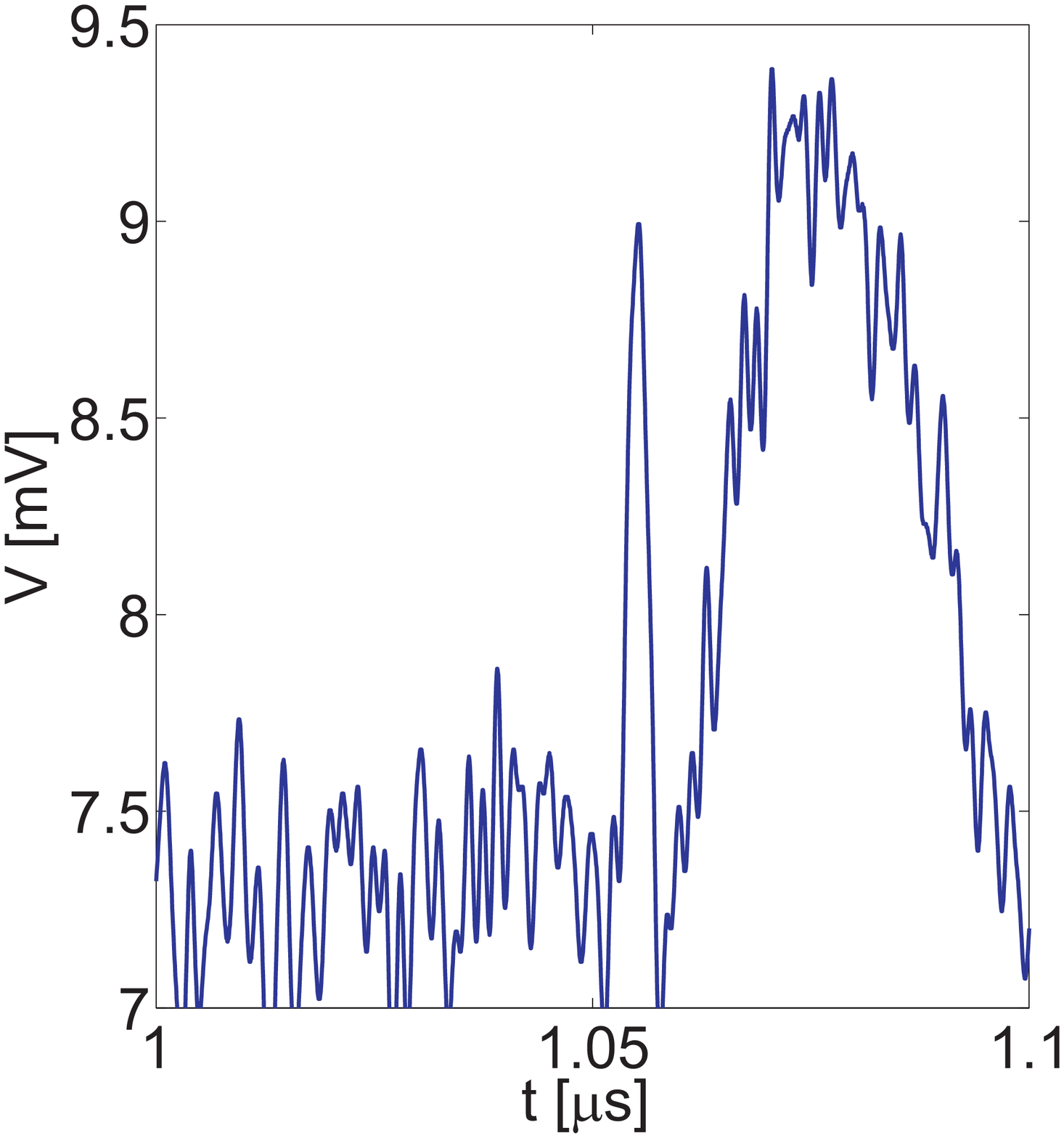}
 \includegraphics[width=0.4\textwidth]{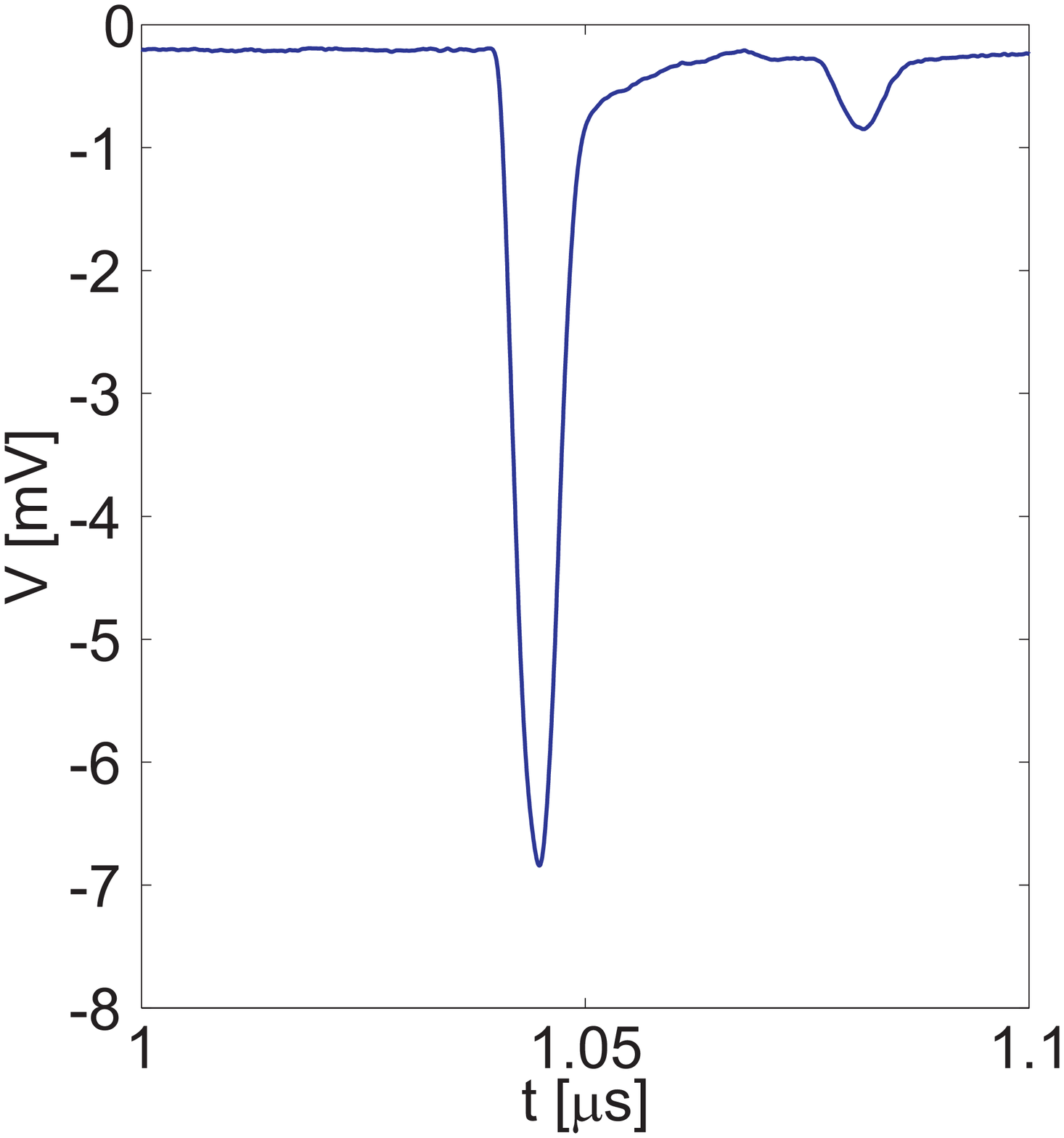}
 \caption{\label{fig:5ns}
(a) Trace obtained for a 5 ns long flux pulse. Only one peak is now observed. (b) The 5 ns pulse. The pulse is limited by the rise time of the pulse generator.}
\end{figure}

By shortening the duration of the flux pulse, a lower limit on the tuning speed can be obtained. In order to observe oscillations for a flux pulse of length $\tau$  the detuning must be sufficiently large so that $2\pi/|\omega_n-\omega_d| \le \tau$, therefore a large amplitude has to be used for the short pulses. figure\,\ref{fig:10ns}(a) shows the obtained trace for a 10 ns long pulse flux (showed in figure\,\ref{fig:10ns}(b)), three peaks are observed with a frequency of $\sim$330\,MHz. There are also some slower oscillations observed after the peaks, these structures are due to reflections of the pulse in the cables. If the pulse is decreased further down to 5 ns one peak can still be observed \ref{fig:5ns}(a). On this sort time scale the pulse starts to be limited by the rise time of the pulse generator \ref{fig:5ns}(b), which is about 2.5 ns. From these measurements it can be concluded that the resonator can be tuned several hundred MHz on a time scale of a few ns.

\section{Conclusions}
In conclusion, we have designed and measured several tunable superconducting
CPW resonators.  We have demonstrated a tunability of more than 700\,MHz for a 4.9
GHz device.  As a figure of merit, we see that we can detune the
devices more than 250 corrected linewidths.  We have also demonstrated
that our device can be tuned substantially faster than its decay time,
allowing us to change the frequency of the energy stored in the
cavity.  Having done this in the few photon limit, we therefore assert
that we can tune the frequency of individual microwave photons stored
in the cavity.  This can be done by several hundred MHz on the
timescale of nanoseconds.

\section{Acknowledgments}
The samples were made at the MC2 clean room at Chalmers. This work was supported by the Swedish SSF and VR, and by the Wallenberg foundation.

\bibliographystyle{apsrev}
\bibliography{ref}
\end{document}